\documentclass[aps,pra,twocolumn,showpacs,groupedaddress]{revtex4}  
\usepackage{graphicx}  % needed for figures
\usepackage{dcolumn}   % needed for some tables
\usepackage{bm}        % for math
\usepackage{amssymb}   % for math
\usepackage{amsmath}
\usepackage{float}
\usepackage{enumerate}
\begin{document}

\title{Bi-$\cal{PT}$ symmetry in nonlinearly damped dynamical systems and \\ tailoring $\cal{PT}$ regions with position dependent loss-gain profiles}
\author{S. Karthiga$^{1}$, V.K. Chandrasekar$^{2}$, M. Senthilvelan$^{1}$, M. Lakshmanan$^{1}$}
\address{$^1$ Centre for Nonlinear Dynamics, School of Physics, Bharathidasan University, Tiruchirappalli - 620 024, Tamil Nadu, India.\\
$^2$ Centre for Nonlinear Science \& Engineering, School of Electrical \& Electronics Engineering, SASTRA University, Thanjavur -613 401, Tamil Nadu, India.}
 \begin{abstract}
\par We investigate the remarkable role of position dependent damping in determining the parametric regions of symmetry breaking in nonlinear $\cal{PT}$-symmetric systems. We illustrate the nature of $\cal{PT}$-symmetry preservation and breaking with reference to a remarkable integrable scalar nonlinear system. In the two dimensional cases of such position dependent damped systems, we unveil the existence of a class of novel bi-$\cal{PT}$-symmetric systems which have two fold $\cal{PT}$ symmetries.  We analyze the dynamics of these systems and show how symmetry breaking occurs, that is whether the symmetry breaking of the two $\cal{PT}$ symmetries occurs in pair or occurs one by one.  The addition of linear damping in these nonlinearly damped systems induces competition between the two types of damping.  This competition results in a $\cal{PT}$ phase transition in which the $\cal{PT}$ symmetry is broken for lower loss/gain strength and is restored by increasing the loss/gain strength. We also show that by properly designing the form of the position dependent damping, we can tailor the $\cal{PT}$-symmetric regions of the system.
\end{abstract}
\maketitle
\section{Introduction}
\par In recent times considerable interest has been shown in investigating systems which do not show parity ($\cal{P}$) and time reversal ($\cal{T}$) symmetries separately but which exhibit a combined $\cal{PT}$ symmetry.  These $\cal{PT}$-symmetric systems have several intriguing features such as power oscillations \cite{b1}, absorption enhanced transmission \cite{k2}, double refraction, and non-reciprocity of light propagation \cite{b1}. Thus, these systems open up novel applications in  optics \cite{b1}, quantum optics \cite{c1, c2}, solid state physics \cite{b3}, metamaterials \cite{b4,b5}, optomechanical systems \cite{ar,ar2}, etc. The understanding of $\cal{PT}$-symmetric systems as non-isolated systems with balanced loss and gain has led to the exploration of these systems in mechanics as well as in electronics.  Such observations of $\cal{PT}$-symmetric mechanical and electronic systems provide the simplest ground to experiment on these $\cal{PT}$-symmetric systems \cite{b6, b8, b9, b99, b10}. 
\par {\it{A.Bi-\cal{PT} symmetry}} The above oscillator based $\cal{PT}$-symmetric systems are generically constructed by coupling an oscillator with linear loss to an oscillator with equal amount of linear gain \cite{b8, b9, b99, b10}.  Apart from the above type of systems, there exists a class of interesting dynamical systems with position dependent damping (or position dependent loss-gain profile) where the amount of damping depends on its displacement.  Consequently one can have $\cal{PT}$-symmetric systems even with a single degree of freedom.  In this case, the systems are invariant with respect to the $\cal{PT}$ operation defined by $\cal{P}$: $x \rightarrow -x$, $\cal{T}$: $t \rightarrow -t$, so that $\cal{PT}$: $x \rightarrow -x$, $t \rightarrow -t$ which we denote as the $\cal{PT}$$-1$ operation.  As the position dependent damping term is found to be a nonlinear term in the evolution equation, we call this damping as {\it nonlinear damping} for simplicity.  The main aim of this paper is to investigate the dynamics and underlying novel structures in these systems in comparison with the standard ones.

\par The recent explorations on the damping in systems with one or more atomic-scale dimensions have unveiled that the damping present in these systems is strongly position dependent \cite{k4, k5, acs}.   Ref. \cite{k4} shows that this type of damping in mechanical resonators enhances the figure of merit of the system tremendously. In particular, with this type of damping, a quality factor of $100,000$ has been achieved with graphene resonators.   In addition, such systems are found to play an important role in many areas of physics, biology and engineering \cite{ando} and they are typically called Li\'enard systems or Li\'enard oscillators.   Recently, a class of chemical and biochemical oscillations which are governed by two-variable kinetic equations are shown to be reducible to Li\'enard systems by linear transformations.   As the nonlinear damping term in the Li\'enard systems can act as a damping term or a pumping term depending on the amplitude of the oscillation, through an internal energy source, it gives rise to self sustained oscillations.  The above property enables one to understand and to control several chemical and biochemical oscillations which are discussed in \cite{chem}.   Li\'enard systems are also found to be paradigmatic models in the biological regulatory systems \cite{bio}.  For example, they have been used to model the heart and respiratory systems (van-der Pol equation \cite{vdp1,vdp2}) and the nerve impulse (FitzHugh-Nagumo equations \cite{fitz}).   The Li\'enard equation with a cubic polynomial potential has been used to describe the isotropic turbulence \cite{flu}.  One can also find the appearance of these systems in reaction-diffusion systems \cite{diffu}.
\par Concerning the importance of the above type of nonlinearly damped systems, we here focus on the $\cal{PT}$-symmetric cases of this category.  The Hamiltonian structure \cite{b14} and quantization \cite{b15, bag} of some of the nonlinearly damped $\cal{PT}$-symmetric systems with single degree of freedom have been studied recently, which show interesting symmetry breaking in these systems (see also Section \ref{meeb} below).
\par A proper coupling of two scalar nonlinearly damped $\cal{PT}$-symmetric systems can yield novel bi-$\cal{PT}$-symmetric systems which are invariant with respect to the $\cal{PT}$$-1$ ($x \rightarrow -x$, $y \rightarrow -y$, $t \rightarrow -t$) operation as well as with the $\cal{PT}$$-2$ operation which is defined as $\cal{PT}$$-2$: $x \rightarrow -y$, $y \rightarrow -x$, $t \rightarrow -t$. { Such type of studies on the systems with multiple $\cal{PT}$ symmetries is interesting, for example one can see an earlier paper on such multiple $\cal{PT}$ symmetric cases \cite{newr2}}. In this paper, we point out that the study of bi-$\cal{PT}$ symmetries in such coupled nonlinear damped systems can lead to interesting novel dynamical states of $\cal{PT}$ symmetry preserving and breaking types, besides oscillation death and bistable states. 

\par{ \it{B. Spontaneous symmetry breaking:}} An interesting mechanism that is found to arise in the $\cal{PT}$ symmetric systems is the spontaneous symmetry breaking, where the system in the symmetric state transits to an asymmetric state by the variation of certain parameters.   In classical systems, the simplest state of broken symmetry is the equilibrium state which may correspond to the minimum of the potential but which does not possess all the symmetries underlying the dynamical equation.   Let $G$ be the transformation under which the dynamical equation is invariant.  Then a symmetric state $u=u_s$ corresponds to the state which  remains invariant under the transformation $u_s=G u_s$.  But an asymmetric or symmetry broken state $u_a$ (that may also correspond to the minimum of the potential) is the one that gets transformed into another asymmetric state $u_i=G u_a$ under the transformation $G$.  Here the transformed state $u_i$ also corresponds to an equilibrium of the system. A typical example is the reflection symmetry in a double well quartic anharmonic oscillator.  From a dynamical point of view the spontaneous breaking of symmetries is also manifested in the stability nature of the fixed points and the trajectories around it in the phase space and nature of bifurcations as a system parameter is varied, again as in the case of the double well quartic oscillator undergoing spontaneous $\cal{P}$-symmetry breaking.  In this paper, we also show that the above existence of symmetry preserving/breaking equilibrium states can be identified with the existence or nonexistence of the general solution of the initial value problem underlying the dynamical system satisfying the symmetry and the system can admit more general classes of solution corresponding to symmetry preservation/breaking. 
\par   A universal feature of the standard $\cal{PT}$-symmetric systems is that the $\cal{PT}$ symmetry is broken by increasing the loss/gain strength and is restored by reducing it \cite{b8,b9}. In contrast to this behavior, Liang {\it{et al}}. \cite{k3} have observed a reverse $\cal{PT}$ phase transition phenomenon in a lattice model known as $\cal{PT}$-symmetric Aubry-Andre model \cite{aub}, in which the $\cal{PT}$ symmetry is broken for lower loss/gain strength and is restored for higher loss/gain strength.  They observed this phenomenon only when two lattice potentials that introduce loss/gain in the system are applied simultaneously (which is not observed when a single lattice potential is present).  This type of inverse $\cal{PT}$ phase transition arises as a result of the competition between the two lattice potentials.  Similarly, Miroshnichenko {\it et al}. \cite{Isr} have studied the competing effect of linear and nonlinear loss-gain profile in discrete nonlinear Schr\"{o}dinger system.  { The observation of $\cal{PT}$ restoration at higher loss-gain strengths also attracted wide interests and the recent studies show that it could happen even through an interplay of kinematical and dynamical nonlocalities \cite{newr}.}

\par { \it{C. Nonlinear damping and $\cal{PT}$ symmetry:}} From a different point of view, in the present work, we add a linear damping in addition to the nonlinear damping and study the competing effects of the linear and nonlinear damping forces. With a single nonlinear damping, our system shows $\cal{PT}$ symmetry breaking like the standard $\cal{PT}$-symmetric systems, but as soon we add the linear damping to the nonlinear damping, we observe $\cal{PT}$ restoration at higher loss/gain strength similar to the case of Aubry-Andre model.   Importantly, we illustrate that this competition among the damping terms in addition to the position dependent nature of damping aid in tailoring the $\cal{PT}$ regions of the system.  

\par The organization of the paper is as follows, in section \ref{snd}, we discuss the loss-gain profiles of the scalar $\cal{PT}$-symmetric and non-$\cal{PT}$-symmetric nonlinearly damped systems.  In section \ref{meeb}, we consider a specific model of scalar $\cal{PT}$ symmetric nonlinear damped oscillator, namely the modified Emden equation.  Analyzing the initial value problem of an integrable case explicitly, we greatly clarify the nature of $\cal{PT}$ symmetry preservation/breaking. In section \ref{sbi}, we consider a coupled system with a simple nonlinear damping $h(x,\dot{x})=x\dot{x}$, which is also a bi-$\cal{PT}$-symmetric system.  In section \ref{slin}, in addition to the nonlinear damping, we introduce a linear damping in the system and show the occurrence of $\cal{PT}$ restoration at higher values of loss/gain strength.  In section \ref{gene}, we consider a general coupled system with linear and nonlinear damping and show the tailoring of $\cal{PT}$ regions in the system.  In section \ref{conc}, we summarize the results of our work.  In Appendix \ref{appe1}, we consider the initial value problem of a double-well oscillator and discuss the spontaneous $\cal{P}$-symmetry breaking from solution point of view.  In Appendix \ref{app2} we consider non-$\cal{PT}$ symmetric scalar systems.  In Appendices \ref{apeg1}, \ref{apeg2} and \ref{new}, we have presented the eigenvalues obtained through the linear stability analysis for the systems we considered.

\section{\label{snd} Nonlinearly damped systems-revisited}
 \par To start with, we analyze the loss-gain profiles of position dependent scalar nonlinearly damped systems.  For this purpose, we first consider a system which is described by the second order nonlinear differential equation
\begin{eqnarray}
\ddot{x}+h(x,\dot{x})+g(x)=0. \qquad \qquad  \left(\;\dot{}= \frac{d}{dt}\right)
\label{sec_ord}
\end{eqnarray}
Here, $h(x,\dot{x})=f(x)\dot{x}$ is the position dependent damping which we call for simplicity as the nonlinear damping term.  Also, $f(x)$ is taken as a non-constant function in $x$.  The above equation can be considered as a dynamical system on its own merit, often with a nonstandard Hamiltonian description \cite{b14}, or as a conservative nonlinear oscillator perturbed by a nonlinear damping force $h(x,\dot{x})$ which supplies or absorbs energy at different points in the $(x, \dot{x})$ phase space,
\begin{eqnarray}
\ddot{x}+g(x)=-h(x,\dot{x})=-f(x) \dot{x}.
\label{sec_ord2}
\end{eqnarray}
The kinetic and the potential energies of the unperturbed particle are given respectively by
\begin{eqnarray}
T(\dot{x})=\frac{1}{2}\dot{x}^2; \qquad V(x)=\int g(x) dx.
\label{pot}
\end{eqnarray}
Thus the total energy of the particle in the potential $V(x)$ when $h(x, \dot{x})=0$ is
\begin{eqnarray}
E= \frac{1}{2}\dot{x}^2+\int g(x) dx.
\label{egy}
\end{eqnarray}
The rate of change of energy of the particle is
\begin{eqnarray}
\frac{dE}{dt}=\dot{x}(\ddot{x}+g(x)).
\label{de}
\end{eqnarray}
From Eq. (\ref{sec_ord}), we can write
\begin{eqnarray}
\frac{dE}{dt}=-\dot{x}h(x, \dot{x})=-f(x) \dot{x}^2.
\label{deh}
\end{eqnarray}
\begin{figure}[htb!]
\begin{center}
   \includegraphics[width=1.0\linewidth]{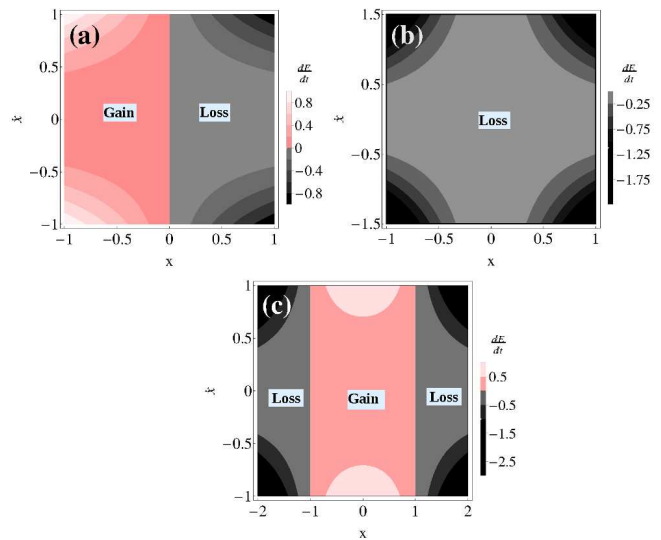}
\end{center}
   \caption{(Color online) Loss-gain profiles $\frac{dE}{dt}$ of the systems given by (a) Eq. (\ref{pt}), (b) Eq. (\ref{npt}) and (c) Eq. (\ref{lim}) in the $(x,\dot{x})$ space:  The pink shaded regions in the figures correspond to the regions in which $\frac{dE}{dt}$ is positive (or it denotes the region in which gain is present).  Similarly, the gray shaded regions denote the regions in which $\frac{dE}{dt}$ is negative.}
\label{den}
   \end{figure}
\par If the quantity $\frac{dE}{dt}<0$ (or $\dot{x} h(x, \dot{x}) >0$) in a region in $(x ,\dot{x})$ phase space, then the energy is withdrawn from the system for the states lying in this region and the role of $h(x, \dot{x})$ is like a damping or loss term and if $\frac{dE}{dt}>0$ (or $\dot{x} h(x, \dot{x}) <0$), then in the corresponding region the effect of $h(x, \dot{x})$ is like negative-damping or gain.

\par The above type of nonlinearly damped systems can be classified as $(i)$ $\cal{PT}$-symmetric systems and $(ii)$ non-$\cal{PT}$-symmetric systems depending on the form of $h(x, \dot{x})$, whereas all linearly damped systems are always non-$\cal{PT}$-symmetric. Here, the $\cal{PT}$-symmetric systems  are  those systems that are invariant under the combined operation of $\cal{PT}$ (and not individual operation of $\cal{P}$ or $\cal{T}$): $x \rightarrow -x$, $t \rightarrow -t$. We denote this as $\cal{PT}$$-1$ symmetry (in order to distinguish it from the additional $\cal{PT}$ symmetry in two dimensional systems).  Then $\cal{PT}$$-1$ symmetric systems belonging to (\ref{sec_ord}) are those systems where $h(x, \dot{x})$ is a nonlinear function in $x$, $\dot{x}$ that is odd in $x$ as well as $\dot{x}$.  In this article, we focus our attention towards the systems with $h(x, \dot{x})=f(x) \dot{x}$, where $f(x)$ and $g(x)$ in (\ref{sec_ord}) are odd functions.  Systems of the form (\ref{sec_ord}) which do not meet this requirement are non-$\cal{PT}$-symmetric.    These non-$\cal{PT}$-symmetric systems are typically of two types, $(i)$ systems exhibiting damped oscillations and $(ii)$ systems admitting limit cycle oscillations.  In the following we present specific examples of these three cases:

\begin{enumerate}
\item $\cal{PT}$-symmetric conservative system - Modified Emden Equation (MEE)\cite{b14, gen}:
\begin{eqnarray} \ddot{x}+\alpha x\dot{x}+\beta x^3+\omega_0^2 x&=&0  \label{pt} \end{eqnarray}
\item Non-$\cal{PT}$-symmetric damped system\cite{damp}:
\begin{eqnarray}\ddot{x}+\alpha x^2 \dot{x}+\beta x^3+\omega_0^2 x&=&0 \label{npt}  \end{eqnarray}
\item Limit cycle oscillator (van der Pol oscillator) \cite{bok}:
\begin{eqnarray}\ddot{x}+(x^2-1)\dot{x}+\omega_0^2 x&=&0. 
\label{lim}
\end{eqnarray} 
\end{enumerate}

\par The system (\ref{pt}) is known as the modified Emden equation and is obviously invariant under the $\cal{PT}$$-1$ operation.  The $\cal{PT}$-symmetric nature of this system \cite{b14} and its quantization \cite{b15} have been studied for the specific case $\beta=\frac{\alpha^2}{9}$ which admits symmetry breaking states for $\lambda<0$.  A critical analysis of the $\cal{PT}$- symmetry of (\ref{pt}) is given in section \ref{meeb}.  The systems given in Eqs. (\ref{npt}) and (\ref{lim}) are examples of non-$\cal{PT}$-symmetric ones, as the damping term in these cases are found to be even functions of $x$.  The system (\ref{npt}) admits damped oscillations, while the system (\ref{lim}) (the famous van der Pol oscillator) is found to have self sustained oscillations which is also noted in Appendix \ref{app2}. 
\par Figure \ref{den} shows the loss-gain profiles corresponding to Eqs. (\ref{pt}), (\ref{npt}) and (\ref{lim}), which are obtained by substituting the corresponding forms of $f(x)$ in Eq. (\ref{deh}) . From the loss-gain profile (shown in Fig.\ref{den}(a)) corresponding to the $\cal{PT}$$-1$ symmetric case (\ref{pt}), we can find that we have varying loss along the positive $x-$ axis and varying gain along the negative $x-$ axis. The amount of gain present for $x<0$ is balanced by the amount of loss present for $x>0$. Then from Figs. \ref{den}(b) and \ref{den}(c), we can see that in the case of non-$\cal{PT}$-symmetric systems, the loss and gain will not be balanced.  In the case of the non-$\cal{PT}$ damped oscillator (\ref{npt}), from Fig. \ref{den}(b) we can find that loss is present everywhere in space. In the case of limit cycle oscillator (\ref{lim}), from Fig. \ref{den}(c), we can find that gain exists in the region $|x|<1$ and loss exists in the region $|x|>1$.  This clearly shows that in this case, the amount of loss present in the $(x, \dot{x})$ space is not balanced by an equal amount of gain.

\begin{figure}[htb!]
\begin{center}
   \includegraphics[width=1.0\linewidth]{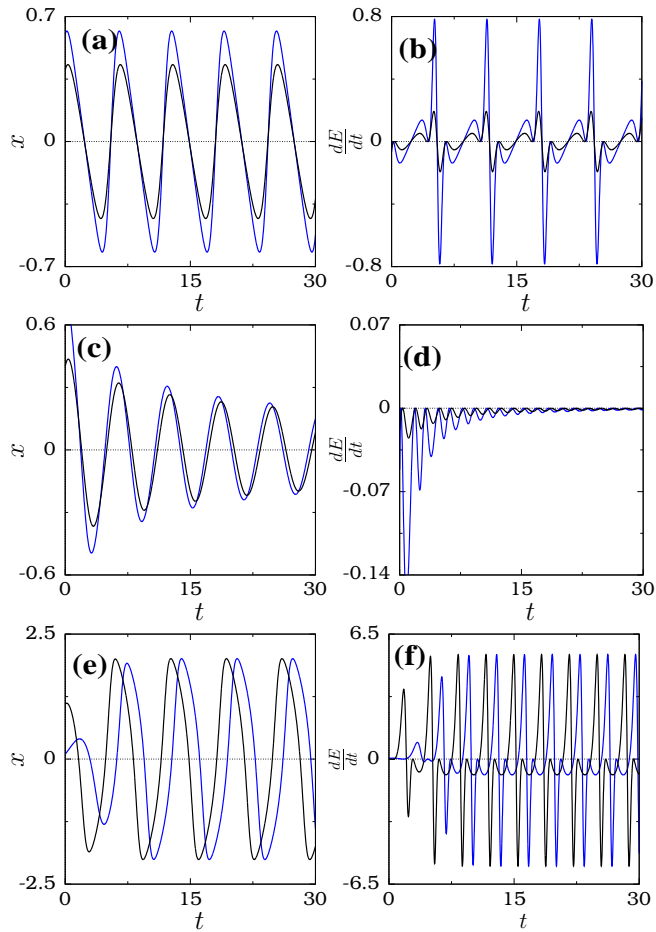}
\end{center}
   \caption{(Color online) Figures $(a)$, $(c)$ and $(e)$ depict the solution of Eqs. (\ref{pt}), (\ref{npt}) and (\ref{lim}), respectively, for two different initial conditions.  Figures $(b)$, $(d)$ and $(f)$ show the corresponding rates $\frac{dE}{dt}$ as a function of time.}
\label{mee}   
\end{figure}  

\par From Fig. \ref{mee}, we can see that in the $\cal{PT}$-symmetric and limit cycle oscillator cases, there exists periodic and self sustained oscillations (Figs. \ref{mee}(a), \ref{mee}(e)), respectively, and in the non-$\cal{PT}$-symmetric damped oscillator case (Fig. \ref{mee}(c)), we have damped oscillations.  The corresponding rates of change of energy $\frac{dE}{dt}$ profiles are shown in Figs. \ref{mee}(b), \ref{mee}(d) and \ref{mee}(f), respectively.

\par Comparing the periodic oscillations (Figs. \ref{mee}(a) and \ref{mee}(e)) corresponding to the $\cal{PT}$-symmetric oscillator case (Eq. (\ref{pt})) and the limit cycle oscillator case (Eq. (\ref{lim})), we can find that the $\cal{PT}$-symmetric system takes up different paths for different initial conditions but the limit cycle oscillator for different initial conditions tends to a particular path as time $t \rightarrow \infty$.   The reason is that the balanced loss-gain profile (shown in Fig. \ref{den}(a)) of the $\cal{PT}$-symmetric system allows it to have multiple paths along which net $\frac{dE}{dt}$ is zero.  But in the case of limit cycle oscillator, Fig. \ref{den}(c) shows that the loss and gain are not balanced in the ($x,\dot{x}$) space.  Thus the paths along which total $\frac{dE}{dt}$ is zero are limited in this case.  Consequently the phase space of limit cycle oscillators contains isolated paths only.
\par Now let us consider a system of coupled nonlinear damped oscillators (for simplicity we consider a linear coupling)
\begin{eqnarray}
\ddot{x}+h_1(x,\dot{x})+h_2(x,\dot{x})+g(x)+\kappa y&=&0, \nonumber \\
\ddot{y}+h_1(y,\dot{y})-h_2(y,\dot{y})+g(y)+\kappa x&=&0,
\label{cop}
\end{eqnarray}
where $h_1(x,\dot{x})= f_1(x)\dot{x}$ and $h_2(x,\dot{x})=f_2(x)\dot{x}$ are the two position dependent nonlinear damping terms.  Here, the functions $f_1(x)$ and $f_2(x)$ are chosen to be odd and even functions in $x$, respectively, and also the function $g(x)$ is chosen as odd.  Consequently, the system becomes symmetric with respect to the $\cal{PT}$$-2$ operation (which is defined as $\cal{PT}$$-2$: $x \rightarrow -y$, $y \rightarrow -x$, $t \rightarrow -t$).   Now, by making $f_2(x)$ to be zero, the system is symmetric with respect to both $\cal{PT}$$-1$ and $\cal{PT}$$-2$ operations. (Here $\cal{PT}$$-1$ corresponds to the operation $x \rightarrow -x$, $y \rightarrow -y$, $t \rightarrow -t$.)  Thus the system is bi-$\cal{PT}$-symmetric in this case.

\par Similar to the scalar case, we can consider the above system as a system of two coupled oscillators
\begin{eqnarray}
\ddot{x}+g(x)+\kappa y&=&0, \nonumber \\
\ddot{y}+g(y)+\kappa x&=&0,
\label{cop2}
\end{eqnarray}
acted upon by additional external forces $h_1(x,\dot{x})$ and $h_2(x,\dot{x})$.  The total energy of the system  (in the absence of nonlinear damping) is given by 
\begin{eqnarray}
E=\frac{1}{2}\dot{x}^2+\int g(x) dx+\frac{1}{2}\dot{y}^2+\int g(y) dy +\kappa x y.
\label{e}
\end{eqnarray}
The rate of change of energy in the system due to the weak nonlinear damping term as specified by Eq. (\ref{cop}) is given by
\begin{eqnarray}
\frac{dE}{dt}=-\dot{x}[h_1(x,\dot{x})+h_2(x,\dot{x})]-\dot{y}[h_1(y,\dot{y})-h_2(y,\dot{y})]. \quad
\label{dee}
\end{eqnarray}
The above expression shows that similar to the scalar case, the coupled system (\ref{cop})  also has position dependent loss-gain profile.  Further, the question whether a nonstandard Hamiltonian description similar to the scalar case (Section \ref{meeb}) exists for (\ref{cop}) has not yet been answered in the literature as far as the knowledge of the authors goes, though a class of such systems has recently been identified \cite{rg,ak}.
\section{\label{meeb} $\cal{PT}$ symmetry breaking in the modified Emden equation}
The system mentioned in Eq. (\ref{pt}), namely
\begin{eqnarray}
\ddot{x}+\alpha x \dot{x}+\beta x^3+ \lambda x=0, \qquad \lambda=\omega_0^2, \label{pt22}
\end{eqnarray}
is the simplest example for $\cal{PT}$$-1$ symmetric system.  The reversible nature of the system has been studied and this equation is used as a normal form for describing the symmetry breaking bifurcation in certain reversible systems which includes an externally injected class B laser system \cite{poli}.  This $x \dot{x}$ type damping has been found to appear in many chemically relevant kinetic equations \cite{chem}. The model is found to be useful in fluid mechanics where the linearly forced isotropic turbulence \cite{flu} can be described in terms of a cubic Li\'enard equation which is of the form similar to (\ref{pt22}).  This system is also found to appear in some important astrophysical phenomena and it occurs in the study of equilibrium configurations of a spherical cloud acting under the mutual attraction of its molecules and is subject to the thermodynamic laws \cite{astro}.  Eq. (\ref{pt22}) is known to admit a nonstandard conservative Hamiltonian description \cite{gen} and interesting dynamical properties \cite{damp}.  In particular, the specific choice $\beta =\frac{\alpha^2}{9}$ admits  isochronous properties \cite{b14} (see below) and can  be even quantized in momentum space, exhibiting $\cal{PT}$ symmetry and broken $\cal{PT}$ symmetry as shown by Chithiika Ruby et al \cite{b15} recently, see subsection \ref{3b} below.
\subsection{\label{3a}Linear stability analysis}
\par  Let us analyze the dynamical behavior of the system (\ref{pt22}) qualitatively through a linear stability analysis.  Eq. (\ref{pt22}) can be rewritten as
\begin{eqnarray}
\dot{x}&=&x_1 \nonumber \\
\dot{x_1}&=&-\alpha x x_1-\beta x^3-\lambda x.
\label{ptf}
\end{eqnarray}
  This system has a trivial equilibrium point $E_0$: $(x^*, x_1^*)=(0,0)$ and a pair of non-trivial equilibrium points symmetrically positioned along $x$-axis about $x=0$, $E_{1,2}$: $(\pm \sqrt{-\frac{\lambda}{\beta}},0)$ (which exist only if $\lambda<0$ or $\beta<0$).  In our following analysis, we take $\beta>0$ and so $E_{1,2}$ exist only for $\lambda<0$.  The Jacobian matrix corresponding to the system (\ref{ptf}) is given by
\begin{eqnarray}
J=\left[ \begin{array}{cc}
0\quad&1\\
-\alpha x_1^*-3\beta x^{*^2}-\omega_0^2\;\;&\; -\alpha x^*
\end{array} \right]
\end{eqnarray}
    The eigenvalues of $J$ corresponding to the equilibrium point $E_0$ are $\mu^{(0)}_{1,2}=$ $\pm i \sqrt{\lambda}$.  Similarly, the eigenvalues of $J$ corresponding to $E_1$ and $E_2$ are  $\mu^{(1)}_{1,2}=\frac{1}{2\sqrt{\beta}}(-\alpha \sqrt{-\lambda} \pm \sqrt{-\lambda (\alpha^2-8 \beta)})$, $\mu^{(2)}_{1,2}=\frac{1}{2\sqrt{\beta}}(\alpha \sqrt{-\lambda} \pm \sqrt{-\lambda (\alpha^2-8 \beta)})$.
\par
\begin{figure}[htb!]
\begin{center}
   \includegraphics[width=8cm]{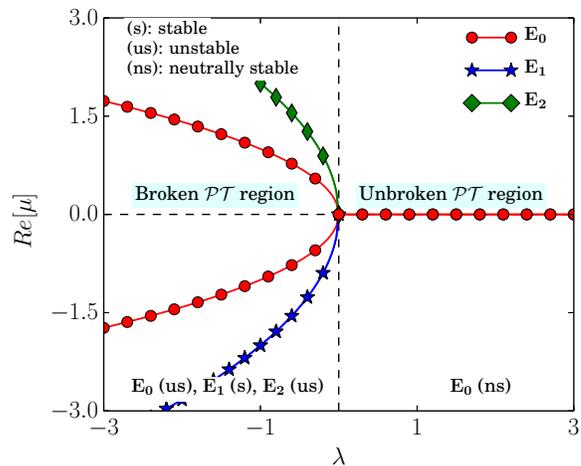}
\end{center}
   \caption{(Color online) Plot of the real part of the eigenvalues of $J$ associated with the equilibrium point $E_0$, $E_1$, $E_2$ of the system (\ref{pt22}) for the values of $\alpha=2$ and $\beta=1$.}
\label{figme}
 \end{figure}

  The real part of the eigenvalues of $J$ associated with the above equilibrium points are given in Fig. \ref{figme}.  The figure shows that in the region $\lambda>0$, the equilibrium point $E_0$ alone exists and all the eigenvalues of $E_0$ are found to be pure imaginary (or $Re[\mu]=0$).  So in the region $\lambda>0$, periodic oscillations exist in the system corresponding to which the phase trajectories around the equilibrium point $E_0$ preserve their structure under $\cal{PT}$ operation, where $E_0$ itself remains invariant: $\cal{PT}$$[E_0]$$=E_0$. Thus $\cal{PT}$-symmetry is unbroken while $\lambda>0$.  But by varying $\lambda$ to $\lambda<0$, a pair of equilibrium points ($E_1$ and $E_2$) with opposite stabilities arise, where $E_1$ is stable (as all the eigenvalues have $Re[\mu] <0$) while $E_2$ is unstable (as all eigenvalues have $Re[\mu] >0$). In this region $E_0$ becomes a saddle (as one of the eigenvalues of $E_0$ has $Re[\mu]>0$ and the other eigenvalue has $Re[\mu]<0$ ).  Under the $\cal{PT}$ operation, $E_1$ gets transformed to $E_2$ and vice-versa: $\cal{PT}$[$E_1$]$=E_2$ and $\cal{PT}$[$E_2$]$=E_1$ so that the $\cal{PT}$ symmetry gets broken.  Correspondingly the trajectories around  $E_1$ get transformed to trajectories around $E_2$ and vice-versa under the $\cal{PT}$ operation. { Note that the above kind of bifurcations fall within the scope of Thom's catastrophe theory \cite{cata}.}
\begin{figure}[htb!]
\begin{center}
   \includegraphics[width=8.0cm]{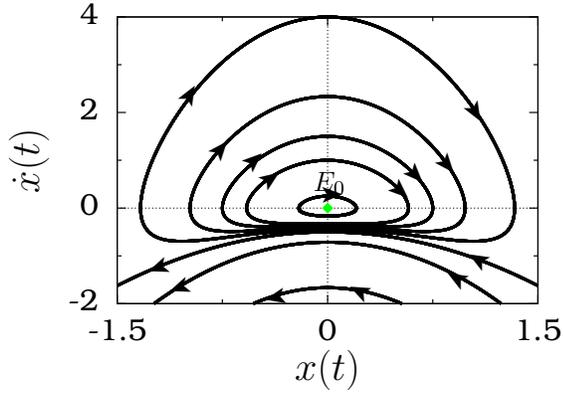}
\end{center}
   \caption{(Color online) Phase portrait of the system (\ref{pt22}) for  $\lambda=1$, $\alpha=3$ and $\beta=1$. The green colored diamond in the figure denotes the position of the neutrally stable equilibrium point $E_0$.}
\label{ph}
 \end{figure}
\begin{figure}[htb!]
\begin{center}
   \includegraphics[width=8.0cm]{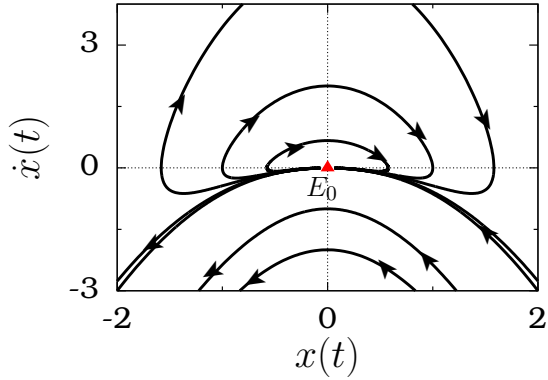}
\end{center}
   \caption{(Color online) Phase portrait of the system (\ref{pt22}) at the bifurcation point  $\lambda=0$ with $\alpha=3$ and $\beta=1$.   }
\label{ph0}
 \end{figure}
\begin{figure}[htb!]
\begin{center}
   \includegraphics[width=8.0cm]{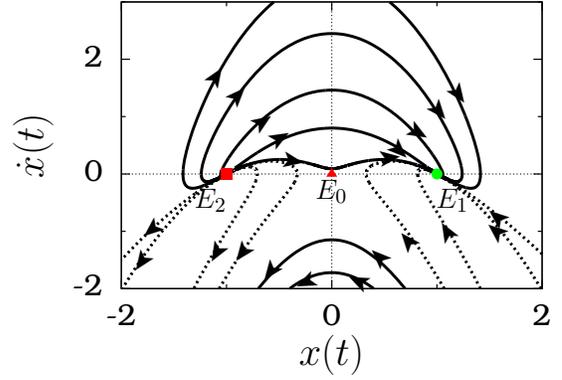}
\end{center}
   \caption{(Color online) Phase portrait of the system (\ref{pt22}) for  $\lambda=-1$, $\alpha=3$ and $\beta=1$. The green circle denotes the stable node type equilibrium point $E_1$, red colored triangle and square correspond to the saddle type equilibrium point ($E_0$) and unstable node type of equilibrium point $E_2$ respectively. The continuous line denotes the orbits corresponding to symmetric solutions and the dashed lines corresponds to that of asymmetric solutions. }
\label{ph2}
 \end{figure}
\par To appreciate these aspects more clearly, we plot the phase portraits of the system for the explicitly integrable case $\beta =\frac{\alpha^2}{9}$, obtained from the exact solutions of the system  \cite{b14}.  A qualitatively similar set of phase portraits results for the general case $\beta \neq \frac{\alpha^2}{9}$, which can be drawn through a numerical analysis.  
\subsection{\label{3b} The exactly integrable case: $\beta=\frac{\alpha^2}{9}$}
\par We consider the specific case $\beta=\frac{\alpha^2}{9}$ of Eq. (\ref{pt22}), namely
\begin{eqnarray}
\ddot{x}+\alpha x \dot{x}+\frac{\alpha^2}{9}x^3+\lambda x=0 
\label{pt23}
\end{eqnarray}
or equivalently 
\begin{eqnarray}
\dot{x}&=&y, \nonumber \\
\dot{y}&=&-\alpha x y - \frac{\alpha^2}{9} x^3 - \lambda x.
\label{fir23}
\end{eqnarray}
Eq. (\ref{pt23}) or (\ref{fir23}) admits a nonstandard Lagrangian/ conservative Hamiltonian description \cite{b14} with
\begin{eqnarray}
L = \frac{27 \lambda^3}{2 \alpha^2}\left( \frac{1}{\alpha \dot{x}+\frac{\alpha^2 }{3} x^2+3 \lambda} \right)+\frac{3 \lambda}{2 \alpha} \dot{x}-\frac{9 \lambda^2}{2 \alpha^2}.
\label{lagr}
\end{eqnarray}
Then the canonically conjugate momentum is 
\begin{eqnarray}
p=-\frac{27 \lambda^3}{2 \alpha}\left(\frac{1}{(\alpha \dot{x}+\frac{\alpha^2 }{3} x^2+3 \lambda)^2} \right)+\frac{3 \lambda}{2 \alpha},
\label{p}
\end{eqnarray}
so that the Hamiltonian $H$
\begin{small}
\begin{eqnarray}
H&=&\frac{9 \lambda^2}{2}\left(\frac{((\dot{x}+\frac{\alpha}{3}x^2)^2+\lambda x^2)}{(\alpha \dot{x}+\frac{\alpha^2 }{3} x^2+3 \lambda)^2}\right) \qquad \qquad \qquad \qquad \nonumber \\
&=&\frac{9 \lambda^2}{2 \alpha^2}\left[2-2\left(1-\frac{2\alpha p}{3 \lambda}\right)^\frac{1}{2}+\frac{\alpha^2 x^2}{9 \lambda}-\frac{2 \alpha p}{3 \lambda}-\frac{2 \alpha^3 x^2 p}{27 \lambda^2}\right]  \qquad
\label{h}
\end{eqnarray}
\end{small}
which is a conserved quantity and we may call it  as the 'energy' $\cal{E}$. 
% Note that the resultant Euler-Lagrange equation from (\ref{lagr}) or the Hamilton's canonical equation from (\ref{h}) is equivalent to (\ref{pt23}) or (\ref{fir23}), respectively.
\par Now the exact solution of (\ref{pt23}) for the three cases $\lambda>0$, $\lambda=0$ and $\lambda<0$ are as follows \cite{b14}:\\

(i) \underline{Case-$1$: $\lambda>0$:} Here one has periodic solutions of (\ref{pt23}) or (\ref{fir23}) as
\begin{subequations}
\label{perme}
\begin{eqnarray}
x(t)&=&\frac{A\sin(\omega_0 t+\delta)}{1-A\frac{\alpha}{3 \omega_0}\cos(\omega_0 t+\delta)},\quad \omega_0=
\sqrt{\lambda} \label{prx}  \\
\dot{x}(t)&=&\frac{A\omega_0 \cos(\omega_0 t+\delta)}{1-A\frac{\alpha}{3 \omega_0}\cos(\omega_0 t+\delta)}-\frac{\alpha}{3}x^2(t),
\label{prxd}
\end{eqnarray}
\end{subequations}
where, $A$ and $\delta$ are constants.  Note that the solution is periodic and bounded for $0 \leq A < \frac{3 \omega_0}{\alpha}$.  For $A \geq \frac{3 \omega_0}{\alpha}$, the solution is singular and periodic.  Also one can evaluate from (\ref{h}) using (\ref{perme}) the 'energy' in this case as
\begin{eqnarray}
H\,=\,{\cal{E}}=\frac{1}{2} \omega_0^2 A^2.
\label{he}
\end{eqnarray}

(ii) \underline{Case-$2$: $\lambda=0$:}  One has a decaying type or front like solution in this case as 
\begin{subequations}
\label{dec0}
\begin{eqnarray}
x(t)&=&\frac{I_1+t}{\frac{\alpha t^2}{6}+\frac{I_1 \alpha t}{3}+I_2}, \qquad  \label{dec01} \\
\mathrm{and} \qquad \quad && \nonumber \\ \quad \dot{x}(t)&=&\frac{1}{\frac{\alpha t^2}{6}+\frac{I_1 \alpha t}{3}+I_2}-\frac{\alpha}{3}x^2(t), \label{dec02}
\end{eqnarray}
\end{subequations}
such that 
\begin{eqnarray}
H={\cal{E}}= 0,
%H={\cal{E}}= \frac{6 I_2 - \alpha I_1}{3 \alpha} 
\label{h0}
\end{eqnarray}
where $I_1$ and $I_2$ are arbitrary constants.\\

(iii) \underline{Case-$3$: $\lambda<0$:} Here we have the general solution
\begin{subequations}
\label{decme}
\begin{eqnarray}
x(t)&=&\frac{3\sqrt{|\lambda|}(I_1 e^{\sqrt{|\lambda|}t}-e^{-\sqrt{|\lambda|}t})}{\alpha(I_1 I_2 +I_1 e^{\sqrt{|\lambda|}t}+e^{-\sqrt{|\lambda|}t})} \label{decm2} \\
\dot{x}(t)&=&\frac{3|\lambda|(I_1 e^{\sqrt{|\lambda|}t}+e^{-\sqrt{|\lambda|}t})}{\alpha(I_1 I_2 +I_1 e^{\sqrt{|\lambda|}t}+e^{-\sqrt{|\lambda|}t})}-\frac{\alpha}{3} x^2(t). \quad \label{decm1}
\end{eqnarray}
\end{subequations}
with 
\begin{eqnarray}
H={\cal{E}}=\frac{18 |\lambda|^2}{\alpha^2}\frac{1}{I_1 I_2^2},
\label{h1}
\end{eqnarray}
where $I_1$ and $I_2$ are arbitrary constants.
\par Now treating the nonlinear differential equation (\ref{pt23}) or (\ref{fir23}) as a dynamical system, we shall consider the solution of its initial value problem (IVP) admitting the $\cal{PT}$ symmetry.  Since we require the $\cal{PT}$ symmetry to be valid for the entire duration of evolution, starting from the initial reference time which may be taken without loss of generality as $t=0$, we require the initial values of the dynamical variables corresponding to a definite 'energy' satisfy the $\cal{PT}$-symmetry conditions ($x(t) \rightarrow -x(-t)$, $t \rightarrow -t$, $\dot{x}(t) \rightarrow \dot{x}(-t)$):
\begin{eqnarray}
x(0)&=&c_1=-x(0)  \nonumber \\
\dot{x}(0)&=& c_2 = \dot{x}(0)
\label{c1x0}
\end{eqnarray}
where $c_1$ and $c_2$ are arbitrary constants.  Then one can identify two possibilities. \\
(i) \underline{$\cal{PT}$-symmetric solution:} 
\begin{eqnarray}
c_1=0, \quad c_2=c
\label{c1c2c}
\end{eqnarray}
such that 
\begin{eqnarray}
{\cal{PT}}[x(t)]&=&-x(-t)=x(t), \nonumber \\
  {\cal{PT}}[\dot{x}(t)]&=&\dot{x}(-t)=\dot{x}(t), \;\; \mathrm{for\;all} \;t \geq 0,
\label{syncond}
\end{eqnarray}
(ii) \underline{$\cal{PT}$-asymmetric solution:} \\
One can consider two distinct values
\begin{eqnarray}
x_1(0)=c_1, \quad x_2(0)=-c_1, \quad c_1 \neq 0
\label{neq}
\end{eqnarray}
such that for $t>0$, one can have a disjoint set of two disconnected solutions/trajectories for a given $\cal{E}$:
\begin{eqnarray}
{\cal{PT}}[x_1(t)]&=&-x_1(-t)=x_2(t) \neq x_1(t), \nonumber \\  {\cal{PT}} [\dot{x}_1(t)]&=&\dot{x}_1(-t)=\dot{x}_2(t) \neq \dot{x}_1(t),
\label{x1x2n}
\end{eqnarray}
and 
\begin{eqnarray}
{\cal{PT}} [x_2(t)]&=&-x_2(-t)=x_1(t) \neq x_2(t), \nonumber \\ {\cal{PT}} [\dot{x}_2(t)]&=&\dot{x}_2(-t)=\dot{x}_1(t) \neq \dot{x}_2(t), \;\mathrm{for\; all} \;\; t \geq 0.\qquad \;\;
\label{x1x2nd}
\end{eqnarray}
associated with the same energy value $\cal{E}$.  Since $x_1(t)$ and $x_2(t)$ correspond to two distinct unconnected trajectories in phase space but with  the same 'energy' value, they represent solutions of broken $\cal{PT}$- symmetry. 
\par We now point out explicitly the above type of solutions for the system (\ref{pt23}) or (\ref{fir23}) in the following, depending on the sign of $\lambda$.  We also demonstrate in Appendix A that a similar type of consideration exists for the $\cal{P}$-symmetric system also, for example in the case of the double well cubic anharmonic oscillator.

\subsection{Observation of symmetry breaking from the solution point of view}

\subsubsection*{Case-1: $\lambda>0$: $\cal{PT}$ invariant solutions}
\par Considering the general solution (\ref{perme}) for $\lambda>0$, without loss of generality we consider the solution of the initial value problem with 
\begin{eqnarray}
x(0)=0, \qquad \dot{x}(0)=B=\frac{3A \omega_0^2}{3\omega_0-A\alpha} 
\label{inper}
\end{eqnarray}
which is itself $\cal{PT}$ invariant. This fixes $\delta=0$ in the solution (\ref{perme}).  Then the resultant general solution (\ref{perme}) with $\delta=0$ of the initial value problem is fully $\cal{PT}$-invariant for all $t \geq 0$, that satisfies (\ref{syncond}). The 'energy' associated with the solution is again $\cal{E}$$=$$\frac{1}{2} \omega_0^2 A^2$ as given in (\ref{he}).   Note that the above solution includes the equilibrium point $E_0=(0,0)$ when $A=0$ with the energy $\cal{E}$ taking the minimum value.  The corresponding phase trajectories in ($x, \dot{x}$) space are plotted in Fig. \ref{ph} which form concentric closed curves around $E_0$ as long as $A<\frac{3 \omega_0}{\alpha}$, so that it is a centre type equilibrium point.  The associated eigenvalues of the equilibrium point $E_0$ are $\pm i \sqrt{\lambda}$ (as shown in Sec. \ref{3a} above).  Note that for $A \geq \frac{3 \omega_0}{\alpha} $, the solution becomes singular at finite times giving rise to open trajectories in the phase space Fig. \ref{ph} but which shall show $\cal{PT}$ symmetry. 
\par One can also observe that the phase trajectories are invariant under time translation.  Consequently, the solution corresponding to any other initial condition obtainable from (\ref{perme}) also follows an identical phase trajectory for a given $A$ and so a given value of 'energy' $\cal{E}$ as it is obtained by a time translation which is an allowed symmetry of the original dynamical system (\ref{pt22}).  Hence these solutions may not be treated as distinct from the one corresponding to (\ref{inper}), if time translation symmetry is also included, along with $\cal{PT}$ symmetry.  Due to the reason, no symmetry breaking asymmetric solution exists here. 
\subsubsection*{Case-2: $\lambda=0$ - Bifurcation point}
\par Here again the solutions of the initial value problem with $x(0)=0$, $\dot{x}(0)=\frac{1}{I_2}$ deduced from (\ref{dec0}) corresponding to $\cal{E}$$=0$, satisfy the $\cal{PT}$ symmetry as shown with the phase trajectories in Fig. \ref{ph0}.
\subsubsection*{Case-3: $\lambda<0$ - $\cal{PT}$ symmetry breaking } 
\par In this case one can identify three distinct classes of solutions from the general solution (\ref{decme}) of (\ref{pt23}) or (\ref{fir23}) for $\lambda<0$, namely $x_0(t)$, $x_1(t)$ and $x_2(t)$.  Among them $x_0(t)$ forms the symmetric solution satisfying (\ref{syncond}) and the set $x(t)=(x_1(t),x_2(t))$ satisfying (\ref{x1x2n}) and (\ref{x1x2nd}) constitutes a spontaneously symmetry breaking set of solutions which are discussed below.

(a) \underline{Symmetric solution:} \\
\par The explicit form of the solution satisfying the initial conditions $x_0(0)=0$, $\dot{x}_0(0)=\mathrm{constant}$ turns out to be the following: 
\begin{eqnarray}
x_0(t)&=&\frac{3\sqrt{|\lambda|}( e^{\sqrt{|\lambda|}t}-e^{-\sqrt{|\lambda|}t})}{\alpha( I_2 + e^{\sqrt{|\lambda|}t}+e^{-\sqrt{|\lambda|}t})} \nonumber \\
\dot{x}_0(t)&=&\frac{3|\lambda|(e^{\sqrt{|\lambda|}t}+e^{-\sqrt{|\lambda|}t})}{\alpha( I_2 + e^{\sqrt{|\lambda|}t}+e^{-\sqrt{|\lambda|}t})}-\frac{\alpha}{3} x_0^2(t),
\label{decx0}
\end{eqnarray}
as can be deduced from the general solution (\ref{decme}).  Here $I_2$ is an arbitrary constant.  Note that the solution (\ref{decx0}) satisfies the $\cal{PT}$ symmetry $\cal{PT}$$(x_0(t), \dot{x}_0(t))$ $=$ $(x_0(t), \dot{x}_0(t))$ and that ($x_0, \dot{x}_0$) $=$ $(0,0)=E_0$ in the limit $I_2 \rightarrow \infty$.  Also, we observe that asymptotically, as $t \rightarrow \infty$, ($x_0(t), \dot{x}_0(t)$) $\underset{t \rightarrow \infty}{\longrightarrow}$ ($\frac{3 \sqrt{|\lambda|}}{\alpha},0)= E_1$.  That is all the nonsingular trajectories approach the fixed point $E_1$, except $E_0$, so that $E_0$ is a saddle. \\

(b) \underline{Asymmetric solution:} \\
\par Next we have the other two distinct solutions which break the $\cal{PT}$ symmetry.  The first one is given by 
\begin{eqnarray}
x_1(t)&=& \frac{3\sqrt{|\lambda|}(I_1 e^{\sqrt{|\lambda|}t}-e^{-\sqrt{|\lambda|}t})}{\alpha( -2 + I_1 e^{\sqrt{|\lambda|}t}+e^{-\sqrt{|\lambda|}t})}, \;\; I_1<0, \nonumber \\
\dot{x}_1(t)&=&\frac{3|\lambda|(I_1 e^{\sqrt{|\lambda|}t}+e^{-\sqrt{|\lambda|}t})}{\alpha( -2 +I_1 e^{\sqrt{|\lambda|}t}+e^{-\sqrt{|\lambda|}t})}-\frac{\alpha}{3} x_1^2(t).
\label{decx1}
\end{eqnarray}
Note that ($x_1(0), \dot{x}_1(0)$) $=(\frac{3 \sqrt{|\lambda|}}{\alpha}, \frac{6|\lambda|}{\alpha (I_1-1)})$ and asymptotically ($x_1(\infty),\dot{x}_1(\infty)$) $=$ $(\frac{3 \sqrt{|\lambda|}}{\alpha}, 0)$ $=$ $E_1$.  Also when $I_1 \rightarrow \infty$, $(x_1(0), \dot{x}_1(0))$ tends to $E_1$. Again all the nonsingular trajectories approach $E_1$ asymptotically.
\par Similarly, we have the other distinct set of trajectories 
\begin{eqnarray}
x_2(t)&=& \frac{3\sqrt{|\lambda|}( e^{\sqrt{|\lambda|}t}- I_1 e^{-\sqrt{|\lambda|}t})}{\alpha( -2 +  e^{\sqrt{|\lambda|}t}+I_1 e^{-\sqrt{|\lambda|}t})}, \;\; I_1 <0, \nonumber \\
\dot{x}_2(t)&=&\frac{3|\lambda|( e^{\sqrt{|\lambda|}t}+I_1e^{-\sqrt{|\lambda|}t})}{\alpha( -2 + e^{\sqrt{|\lambda|}t}+I_1e^{-\sqrt{|\lambda|}t})}-\frac{\alpha}{3} x_2^2(t).
\label{decx2}
\end{eqnarray}
Note that ($x_2(0), \dot{x}_2(0)$) $=(-\frac{3 \sqrt{|\lambda|}}{\alpha}, \frac{6 {|\lambda|}}{\alpha}\frac{1}{I_1-1})$.  In the limit $I_1 \rightarrow -\infty$ this approaches the equilibrium point $E_2=$ $(-\frac{3 \sqrt{|\lambda|}}{\alpha},0)$.  Interestingly, these trajectories (except $E_2$) also approach $E_1$ asymptotically: ($x_2(\infty),\dot{x}_2(\infty)$) $=$ $(\frac{3 \sqrt{|\lambda|}}{\alpha}, 0)$.  Note that in the above each distinct trajectory corresponds to the invariant 'energy' $\cal{E}$ $=\frac{18 |\lambda|^2}{\alpha^2} \frac{1}{I_1 I_2^2}$.  
\par The above facts are illustrated by the corresponding phase trajectories for the case $\lambda<0$ in Fig. \ref{ph2}.  In the case $0 < I_1<\infty$ (but not equal to $1$), the evolution corresponding to $x_1(t)$ and $x_2(t)$ from initial time $t=0$ to $\infty$ lie along the same path, the trajectory corresponding to $x_1(t)$ is found to be a part of the trajectory of $x_2(t)$ $(={\cal{PT}}[x_1(t)])$ (for $I_1>1$) as well as that of $x_0(t)$ or the trajectory corresponding to $x_2(t)$ is found to be a part of the trajectory of $x_1(t)$ $(={\cal{PT}}[x_2(t)])$ for $0<I_1<1$ as well as that of $x_0(t)$. Under time translation these trajectories may be mapped onto each other and so may be considered equivalent to the symmetrical trajectories $x_0(t)$. These are not shown explicitly in Fig. 6.
\par But the most important fact is that for $I_1<0$, $x_1(t)$ and $x_2(t)$ are truely asymmetric and so break the $\cal{PT}$ symmetry. Consequently the solutions $x_1(t)$ and $x_2(t)$ give rise to distinct trajectories, depending on the choices of the arbitrary constants $I_1$ and $I_2$.
\par Thus the above detailed analysis of the completely integrable nonlinear damped system (\ref{pt23}) or (\ref{fir23}) establishes the fact that a necessary and sufficient condition for the preservation of { $\cal{PT}$$-1$ symmetry is the existence of a single fixed point which is of $\cal{PT}$$-1$ invariant center type (that is neturally stable fixed point associated with imaginary eigenvalues of the linearized equation). Note that this requirement demands the existence of a single well potential and rules out cases like three well potential for $\cal{PT}$$-1$ symmetry preservation. Also the origin has to be necessarily the fixed point for $\cal{PT}$$-1$ invariance, $x \rightarrow -x$, $t \rightarrow -t$. } The above requirement allows the existence of $\cal{PT}$-symmetric non-isolated periodic solutions around the fixed point corresponding to concentric closed curves as trajectories as shown in Fig.4. Otherwise the $\cal{PT}$ symmetry is broken as confirmed for the $\lambda<0$ case.  The above discussion also confirms that the existence of $\cal{PT}$ symmetric fixed point and $\cal{PT}$-symmetric solutions near it alone does not imply $\cal{PT}$ symmetry of the full system if the fixed point is not of centre type as seen in the case of $\lambda<0$.  Now we can use the above criteria as the basis for $\cal{PT}$ invariance for our further studies.
\par We also note that the above results hold good for the case of standard Hamiltonian type complex classical $\cal{PT}$ symmetric systems also, where one can find that the symmetry implies $x(t)=-x^*(-t)$ which implies $\mathrm{Re}[x(t)]=x_R(t)=-x_R(-t)$, $\mathrm{Im}[x(t)]=x_I(t)=x_I(-t)$, $\mathrm{Re}[p(t)]=p_R(t)=\dot{x}_R(t)=\dot{x}_R(-t)=p_R(-t)$ and $\mathrm{Im}[p(t)]=p_I(t)=\dot{x}_I(t)=-\dot{x}_I(-t)=-p_I(-t)$. Thus in these cases the $\cal{PT}$ preserving fixed point will be of the form ($x_R(t), x_I(t), p_R(t), p_I(t)$) $=$ $(0, c_1,c_2,0)$, where $c_1$ and $c_2$ are arbitrary constants.  The studies on the classical trajectories of complex $\cal{PT}$ symmetric systems show the existence of regular periodic orbits (possibly with some unbounded orbits) in the unbroken $\cal{PT}$ regions and non-periodic or open and irregular trajectories in the case of broken $\cal{PT}$ regions \cite{bend, nana, anj}.  In addition, in \cite{bend,anj} one can also note that the closed orbits are centered around the $\cal{PT}$ preserving fixed point as discussed above which confirms our results.

\section{\label{sbi} A bi-$\cal{PT}$-symmetric system}
\par As a simple case of the coupled nonlinear damped system (\ref{cop}), we consider a system of coupled modified Emden equations (MEE)
\begin{eqnarray}
\ddot{x}+\alpha x \dot{x}+\beta x^3+\omega_0^2 x+\kappa y=0, \nonumber \\
\ddot{y}+\alpha y \dot{y}+ \beta y^3+ \omega_0^2 y+\kappa x=0. 
\label{bip}
\end{eqnarray}
Here, $\alpha$ is the nonlinear damping coefficient, $\kappa$ is the coupling strength and $\omega_0$ is the natural frequency of the system when $\omega_0^2>0$.  However, we will also consider the case $\omega_0^2<0$ corresponding to the double well potential.  It is obvious that the system (\ref{bip}) admits a bi-$\cal{PT}$ symmetry.  $(i)$ It is invariant under the $\cal{PT}$$-1$ symmetry: $x \rightarrow -x$, $y \rightarrow -y$ and $t \rightarrow -t$.  Eq. (\ref{bip}) is also invariant under $(ii)$ $\cal{PT}$$-2$ symmetry: $x \rightarrow -y$, $y \rightarrow -x$ and $t \rightarrow -t$.  Note that the above two symmetries also imply the symmetry $x(t) \rightarrow y(t)$.

\par Eq (\ref{bip}) can be rewritten as
\begin{eqnarray}
\dot{x}&=&x_1, \nonumber \\
\dot{x_1}&=&-\alpha x x_1-\beta x^3-\omega_0^2 x-\kappa y, \nonumber \\
\dot{y}&=&y_1, \nonumber \\
\dot{y_1}&=&-\alpha y y_1-\beta y^3-\omega_0^2 y-\kappa x.
\label{fit}
\end{eqnarray}
The above set of dynamical equations (\ref{fit}) admit five symmetrical equilibrium points, $e_0$, $e_1$, $e_2$, $e_3$ and $e_4$: 
\begin{enumerate}[(i)]
\item The trivial equilibrium point $e_0$: ($x^*,x_1^*, y^*, y_1^*$) = $(0,0,0,0)$.
\item A symmetric pair of non-zero equilibrium points $e_{1,2}$: ($x^*,x_1^*, y^*, y_1^*$)=$(\pm a^*_1,0,\mp a_1^*,0)$, where $a_1^*= \sqrt{\frac{\kappa- \omega_0^2}{\beta}}$.
\item  Another pair of symmetric non-zero equilibrium points $e_{3,4}$: ($x^*,x_1^*, y^*, y_1^*$) = $(\pm a^*_2,0,\pm a_2^*,0)$, where $a_2^*= \sqrt{\frac{-\kappa- \omega_0^2}{\beta}}$.
\end{enumerate}
Besides the above five fixed points there exist four more asymmetric fixed points which turn out to be unstable in the parametric range of our interest.  So we do not consider them in this paper further.
\subsection{Case: $\Omega=\omega_0^2>0$}
\par In analyzing (\ref{bip}), we first consider the case where $\Omega=\omega_0^2>0$. The existence of the above mentioned equilibrium points in different regions in the parametric space for this case is indicated in Table \ref{tab} (for our further studies we let $\beta>0$ in Eq. (\ref{bip}) or (\ref{fit})).
\par Before entering into the classification of unbroken and broken $\cal{PT}$ regions of the system, we note here that the equilibrium points are also playing a key role in identifying symmetry breaking as shown in the scalar case in the previous section.  In this connection, we classify the $\cal{PT}$$-1$ and $\cal{PT}$$-2$ invaraint fixed points of the system (\ref{bip}), which can be identified by looking for the fixed points which satisfy $\cal{PT}$$-k$$[e_i]$$=$$e_i$, where $k=1,2$ and $i=0,1,2,3, 4$. Using this, one can find that the fixed point $e_0$ alone is $\cal{PT}$$-1$ invariant (that is $\cal{PT}$$-1$$[e_0]$$=$$e_0$), while the three fixed points $e_0$, $e_1$ and $e_2$ are $\cal{PT}$$-2$ invarant and the fixed points $e_3$ and $e_4$ are invariant neither under $\cal{PT}$$-1$ symmetry nor under $\cal{PT}$$-2$ symmetry. 
\par { Generalizing the discussion in the previous section, we can identify the following two criteria on the fixed points of the coupled system of the type (\ref{bip}) or (\ref{fit}) for the invariance of $\cal{PT}$$-1$ and $\cal{PT}$$-2$ symmetries:}
\begin{enumerate}[(i)]
\item For the preservation of $\cal{PT}$$-1$ symmetry again one requires the existence of a single fixed point at the origin which is of neutrally stable type.  The requirement that for $\cal{PT}$$-1$ symmetry $x \rightarrow -x$, $y \rightarrow -y$ $t \rightarrow -t$ demands the exclusion of any other fixed point and that the origin will be the sole fixed point. 
\item For the preservation of $\cal{PT}$$-2$ symmetry which demands $x \rightarrow -y$, $y \rightarrow -x$, $t \rightarrow -t$, the criterion is the existence of one or more fixed points which are all $\cal{PT}$$-2$ invariant out of which atleast one should be neutrally stable type.  For example, in the above system (\ref{fit}) as well as (\ref{bip2}) below besides the origin $e_0$, the fixed points $e_1$ and $e_2$ are also $\cal{PT}$$-2$ invaraint and it is sufficient that atleast one of them is neutrally stable for preservation of $\cal{PT}$$-2$ symmetry (see Figs. \ref{alpha} and \ref{je1} below). A specific case is illustrated in Fig. \ref{ph_c} below. 
 \end{enumerate}
 
\begin{table}[tp]%
\small
\begin{center}
\hspace{0.05\linewidth}\begin{tabular}{|p{0.25\linewidth}|p{0.18\linewidth}|p{0.28\linewidth}|p{0.18\linewidth}|}
\hline
\centering{} &{ \centering {$\kappa < -\omega_0^2$}}&{\centering {$-\omega_0^2 \leq \kappa \leq \omega_0^2$}} & {\centering {$\kappa>\omega_0^2$}}
\vspace{0.2cm}
\\
\hline
{\centering{$\Omega=\omega_0^2>0$}} & {\centering {$e_0$, $e_3$, $e_4$}}& {\centering {$\qquad$ $e_0$}} & {\centering {$e_0$, $e_1$, $e_2$}}
\vspace{0.2cm}
\\
\hline
{\centering{$\Omega=\omega_0^2=0$}} & {\centering {$e_0$, $e_3$, $e_4$}}& {\centering {$\qquad$ $e_0$}} & {\centering {$e_0$, $e_1$, $e_2$}}
\vspace{0.2cm}
\\
\hline
\centering{} & {\centering {$\kappa < \omega_0^2$}}&{\centering{ $\omega_0^2 \leq \kappa \leq -\omega_0^2$}} & {\centering{ $\kappa>-\omega_0^2$}}
\vspace{0.2cm}
\\
\hline
{\centering{$\Omega=\omega_0^2<0$}} & {\centering $e_0$, $e_3$, $e_4$}& {\centering~~~~ $e_0$, $e_1$, $e_2$} & {\centering $e_0$, $e_1$, $e_2$}
\\
\centering{} & {}& {\centering ~~~~~ $e_3$, $e_4$} & {}
\vspace{0.2cm}
\\
\hline
\end{tabular}
\end{center}
\caption{Symmetric equilibrium points of (\ref{fit}) in different regions of the $(\kappa,\omega_0)$ parametric space with $\beta>0$ and $\Omega=\omega_0^2>0$, $\omega_0^2=0$ and $\omega_0^2<0$.}
\label{tab}
\end{table}
\subsubsection{Linear Stability Analysis}
\par Now, to explore the regions in which $\cal{PT}$ symmetries are found to be broken and unbroken,  we first deduce the Jacobian matrix obtained from the linear stability analysis of the above system. It is given by
\begin{small}
\begin{eqnarray}
J=\left[ \begin{array}{cccc}
0&1&0&0\\
c_{21}&-\alpha x^*&-\kappa&0\\
0&0&0&1\\
-\kappa&0&c_{43}&-\alpha y^*\\
\end{array} \right],
\label{jac1}
\end{eqnarray}
\end{small}
where $c_{21}=-\alpha x_1^*-3 \beta {x^*}^2 -\omega_0^2$, $c_{43}=-\alpha y_1^*-3 \beta {y^*}^2-\omega_0^2$ and ($x^*,x^*_1,y^*,y_1^*$) are the equilibrium points of (\ref{fit}). The eigenvalues of the above matrix determine the dynamical behavior of the system in the neighborhood of the equilibrium points qualitatively and the results will be helpful in identifying the broken and unbroken  $\cal{PT}$-symmetric regions of the system.  In the unbroken $\cal{PT}$ region, the trajectories of the system, in addition to the evolution equation,  replicate the full symmetry of the system, while in the symmetry broken region it does not.  In order that the trajectories of the system to be symmetric under $\cal{PT}$ operation, it should have a non-isolated periodic nature (due to the  presence of the time reversal operator $\cal{T}$ in the $\cal{PT}$ operator). Thus, we look for the regions of the system parameters for which the equilibrium point is {\it neutrally stable}, that is the eigenvalues of the Jacobian matrix corresponding to the equilibrium point are pure imaginary. These regions give rise to unbroken $\cal{PT}$-symmetric ranges.  The eigenvalues of the linear stability matrix $J$ corresponding to different equilibrium points of the system are presented in the Appendix \ref{apeg1}, where the ranges of linear stability are also discussed.  

\begin{figure}[ht]
   \includegraphics[width=9.3cm]{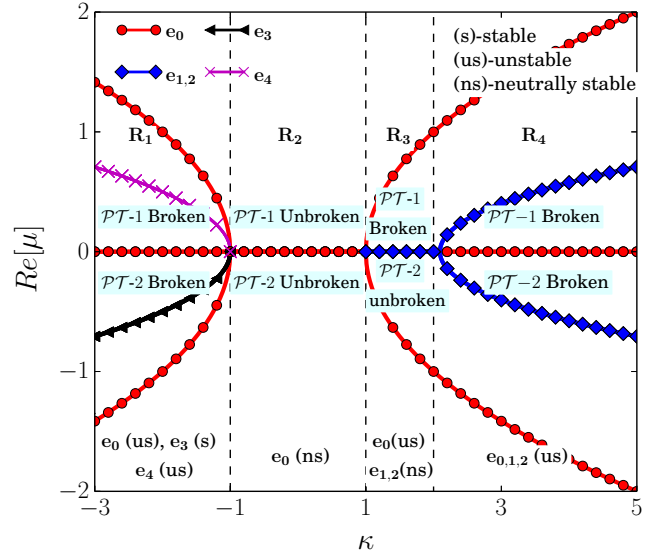}
 \hspace{4cm}  \caption{(Color online) Linear stability of equilibrium points of (\ref{fit}) for $\Omega=\omega_0^2>0$ given in Table. \ref{tab}.  Real parts of eigenvalues of $J$ given by Eq. (\ref{jac1}) are plotted as a function of $\kappa$ for the parameters $\alpha=1.0$, $\beta=1.0$ and $\omega_0=1.0$.}
\label{alpha}   
\end{figure} 
 \par
\begin{figure}[htb!]
\hspace{-0.5cm}
\begin{center}
   \includegraphics[width=9.3cm]{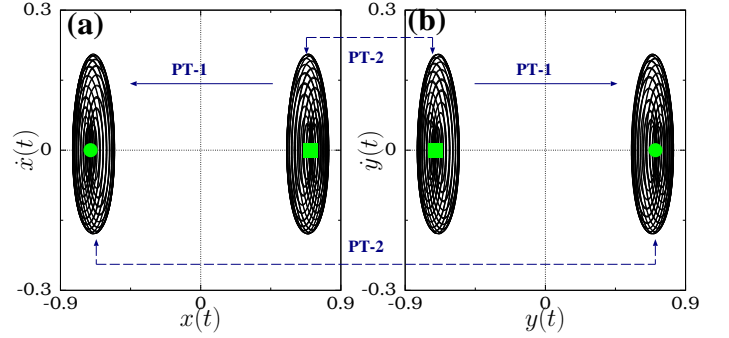}
\end{center}
   \caption{(Color online) Illustration of broken $\cal{PT}$$-1$ and unbroken $\cal{PT}$$-2$ in region $R_3$: (a) ($x -\dot{x}$) (b) ($y - \dot{y}$) projections show the oscillations about the equilibrium points $e_1$ and $e_2$ in the region $R_3$ for $\kappa=1.5$, $\alpha=1.0$, $\beta=1.0$ and $\omega_0^2=1.0$ obtained by solving Eq. (\ref{fit}) numerically (The trajectories away from $e_1$ and $e_2$ are not shown here).  The filled square and the circle represents the position of $e_1$ and $e_2$, respectively. By $\cal{PT}$$-1$ operation on $e_1$ we transit to $e_2$ and so $\cal{PT}$$-1$ symmetry is broken.  But, on the operation of $\cal{PT}$$-2$ on $e_1$ the equilibrium point remains unchanged thereby the symmetry remains unbroken.}
\label{ph_c}   
\end{figure}

 Fixing the parameters $\alpha$, $\omega_0$, $\beta$ as $\alpha=1.0$, $\omega_0=1.0$ and $\beta=1.0$, Fig. \ref{alpha} shows the real parts of the eigenvalues of the equilibrium points $e_0$, $e_{1,2}$ and  $e_{3,4}$ (given in Appendix \ref{apeg1}, Eqs. (\ref{e0bip}), (\ref{e1bip}) and (\ref{e3bip})) under the variation of $\kappa$.  Whenever the real parts of all the eigenvalues of $J$ ($Re[\mu]$) corresponding to an equilibrium point become zero, the eigenvalues are purely imaginary and the latter is said to be neutrally stable.  On the other hand, when all $Re[\mu]$'s corresponding to an equilibrium point are less then zero, it is said to be stable, while the equilibrium point is unstable in all the other cases.   From the forms of the fixed points and the nature of their stability properties, we can identify four separate regions $R_1$, $R_2$, $R_3$ and $R_4$ in the ($\kappa, Re[\mu]$) plane, as follows: 
(i) $R_1$: $\kappa< -\omega_0^2$, (ii) $R_2$: $-\omega_0^2<\kappa< \omega_0^2$,  (iii) $R_3$: $\omega_0^2<\kappa<c \omega_0^2$, where $c$ is given in Eq. (\ref{kapp1}) in Appendix \ref{apeg1}, (iv) $R_4$:  $\kappa> c \omega_0^2$. 
Note that $\omega_0^2=1.0$ in Fig.3.  Then, using the criteria discussed above, we can identify the following facts, as depicted in Fig. \ref{alpha}. 
\begin{itemize}
\item In the region $R_1$ (denoted in Fig. \ref{alpha}), where $\kappa<-\omega_0^2=-1.0$, one can see that three branches appear for $e_0$, and a single branch appears each for $e_3$ and $e_4$. Among the four eigenvalues of $e_0$ (see Eq. (\ref{e0bip})), two are found to be pure imaginary, while the third one has a positive real part and the other has a negative real part.  Thus, in the region $R_1$, there are three branches corresponding to $e_0$.  In each of the cases of $e_3$ and $e_4$, all the eigenvalues have the same real parts (as seen from Eq. (\ref{e3bip})).  Thus, $e_3$ and $e_4$ have a single branch each in Fig. \ref{alpha}.  From the values of $Re[\mu]$ in the region $R_1$, we can find that among the equilibrium points $e_0$, $e_3$ and $e_4$, only $e_3$ is found to be stable.  The stabilization of $e_3$ in the region gives rise to {\it{oscillation death}}.  Here oscillation death in a system of coupled oscillators denotes the stabilization of the system to a non-trivial steady state due to the interaction of oscillators in the system.  We can also note that the equilibrium points $e_3$ and $e_4$ get transformed to one another by both $\cal{PT}$$-1$ and $\cal{PT}$$-2$ operations (that is $\cal{PT}$$-1$[$e_3$]=$e_4$, $\cal{PT}$$-2$[$e_3$]=$e_4$ and vice versa) and the symmetry preserving equilibrium state $e_0$ (that is $\cal{PT}$$-1$[$e_0$]=$e_0$ and $\cal{PT}$$-2$[$e_0$]=$e_0$) is unstable. { Thus both the $\cal{PT}$$-1$ and $\cal{PT}$$-2$ symmetries are broken in this region. }
\item In the region $R_2$, where $-\omega_0^2 \leq \kappa \leq \omega_0^2$ (that is region $-1 \leq \kappa \leq 1$), the equilibrium points $e_3$ and $e_4$ disappear, and $e_0$ alone exists.  The eigenvalues of the equilibrium point $e_0$ in this region are found to be pure imaginary (see also Eq. (\ref{e0bip})).   The neutral stability of the symmetric state $e_0$ signals that in this region $R_2$ both the $\cal{PT}$$-1$ and $\cal{PT}$$-2$ symmetries are unbroken.
\item For $\kappa>\omega_0^2=1$, in the region $R_3$, (defined by Eq. (\ref{kapp1})), $e_0$ loses its stability and gives rise to two new equilibrium points $e_1$ and $e_2$.  These new equilibrium points are found to be neutrally stable.  Further, they also get transformed to each other by $\cal{PT}$$-1$ operation: $\cal{PT}$$-1$[$e_1$] $\Rightarrow$ $\cal{PT}$$-1$[($a_1^*,0,-a_1^*,0$)]=($-a_1^*,0,a_1^*,0$) $=$ $e_2$ and similarly $\cal{PT}$$-1$[$e_2$] $=$ $e_1$.  However, the equilibrium points show invariance under $\cal{PT}$$-2$ operation: $\cal{PT}$$-2$[$e_1$] $\Rightarrow$  $\cal{PT}$$-2$[($a_1^*,0,-a_1^*,0$)]=($a_1^*,0,-a_1^*,0$)= $e_1$ and similarly $\cal{PT}$$-2$[$e_2$] $=$ $e_2$.  The invariance of the equilibrium points $e_1$ and $e_2$ with $\cal{PT}$$-2$ operation is also illustrated in terms of the phase portraits in Fig. \ref{ph_c} obtained by numerical analysis of (\ref{fit}). { As the fixed point preserving $\cal{PT}$$-1$ symmetry ($e_0$) is not of neutrally stable type and due to the coexistence of $\cal{PT}$$-1$ violating fixed points $e_1$ and $e_2$, the $\cal{PT}$$-1$ symmetry is broken in the region. In the case of $\cal{PT}$$-2$ symmetry, all the fixed points ($e_0$, $e_1$ and $e_2$) preserve the symmetry and also two of them ($e_1$ and $e_2$) are neutrally stable.  Thus the $\cal{PT}$$-2$ symmetry is unbroken, as demonstrated in Fig. \ref{ph_c}.} 
  \item For values of $\kappa$ in the region $R_4$ (beyond $R_3$), all the equilibrium points $e_0$, $e_1$ and $e_2$ are found to be unstable.  Thus, both the $\cal{PT}$$-1$ and $\cal{PT}$$-2$ symmetries are found to be broken in the region.
\end{itemize}

\begin{figure}[htb!]
\begin{center}
   \includegraphics[width=0.9\linewidth]{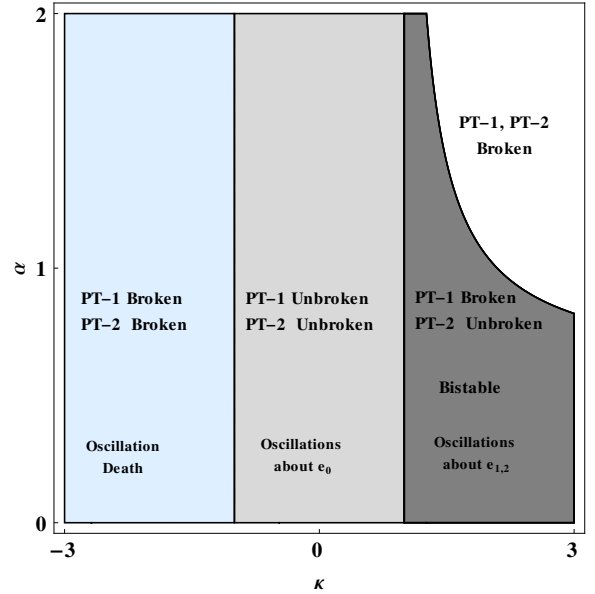}
\end{center}
   \caption{(Color online) Unbroken and broken $\cal{PT}$ regions in the parametric space of ($\kappa,\alpha$) for $\Omega=\omega_0^2>0=1.0$ and $\beta=1.0$. Here the light-gray shaded region denotes the region where both $\cal{PT}$ symmetries are unbroken and the dark-gray shaded region denotes unbroken $\cal{PT}$$-2$ symmetric region. The dark gray shaded regions are denoted as bistable regions in the sense that the equilibrium points $e_1$ and $e_2$ are neutrally stable in that region.  The light blue shaded regions correspond to the oscillation death regions.}
\label{alp}   
\end{figure}
\subsubsection{Dynamics in the $(\kappa, \alpha)$ parametric space}
\par Next, we extend our study as a function of the damping parameter $\alpha$ also.  Fig. \ref{alp} shows the broken and unbroken $\cal{PT}$-symmetric regions corresponding to system (\ref{bip}) in the $(\kappa, \alpha)$ parametric space.  It shows that oscillation death appears in the region $\kappa<-1$ due to the stabilization of $e_3$ as seen earlier in Fig. \ref{alpha} (as can be seen from Eq. (\ref{e3bip}) in Appendix \ref{apeg1}).  Looking at the region $-1 \leq \kappa \leq 1$ in Fig. \ref{alp}, we can observe that the coupled nonlinearly damped system (\ref{bip}) like the scalar case (\ref{pt}) (see Sec. \ref{meeb}), does not show any symmetry breaking on increasing $\alpha$ (see Eq. (\ref{e0bip}) in Appendix \ref{apeg1}).  This is in contrast to the systems with linear damping which show symmetry breaking when the loss/gain strength is increased \cite{b8}. As mentioned in the previous subsection, in this region (that is the region $R_2$ seen in Fig. \ref{alpha}), both $\cal{PT}$$-1$ and $\cal{PT}$$-2$ symmetries are unbroken.  Increasing $\kappa$ further ($\kappa>1$), the system shows breaking of $\cal{PT}$$-1$  symmetry  (for the values of $\kappa>1$ or in the region $R_3$ in Fig. \ref{alpha}) through a pitchfork bifurcation.  In this region, $\cal{PT}$$-2$ symmetry alone is unbroken.  Fig. \ref{alp} shows that the $\cal{PT}$$-2$ symmetry is unbroken only if $\alpha$ is small (from Eq. (\ref{c1bip}) in Appendix \ref{apeg1}) and it is broken for increased $\alpha$ (Note that this type of symmetry breaking at higher values of loss/gain strength is a universal feature of all the $\cal{PT}$-symmetric systems \cite{b8}). On further increasing $\kappa$, Fig. \ref{alp} shows that the $\cal{PT}$ regions with respect to $\alpha$ get reduced. 
\subsubsection{Rotating wave approximation} 
\par In this section, we analyze the stability of the symmetric orbits centered around $e_0$ in the region $R_2$ using the well known rotating wave approximation.  We consider periodic solutions for the system in the region $R_2$ to be of the form   
\begin{eqnarray}
x(t)&=&R_1(t)e^{i{\omega} t}+R^*_1(t)e^{-i{\omega} t}, \nonumber \\ 
y(t)&=&R_2(t)e^{i{\omega} t}+R^*_2(t)e^{-i{\omega} t},
\label{xy}
\end{eqnarray}
where $\omega=\omega_0-\Delta \omega$, and $\Delta \omega$ is a small deviation. 
Here, $R_1(t)$ and $R_2(t)$ are the slowly varying amplitudes with respect to a slow time variable. 
Substituting (\ref{xy}) in (\ref{bip}), and by rotating wave approximation, we obtain 
\begin{eqnarray}
\dot{R}_1&=&\frac{1}{2i{\omega} }(-3\beta|R_1|^2 R_1+(\omega^2-\omega_0^2)R_1-\kappa R_2),\\
\dot{R}_2&=&\frac{1}{2i{\omega} }(-3\beta|R_2|^2 R_2+(\omega^2-\omega_0^2)R_2-\kappa R_1).
\label{phi}
\end{eqnarray}
Now, we separate the real and imaginary parts of the equation as $R_1=a_1+ib_1$, $R_2=a_2+ib_2$.  We have steady periodic solutions when $\dot{a}_i=\dot{b}_{i}=0$, $i=1,2$.  Thus, the equilibrium points of the system represent steady periodic solutions.  The system has five symmetric equilibrium points representing symmetric orbits, which are $E_0$:$(0,0,0,0)$, $E_{1,2}$:$(0,\pm b_{11}^*,0, \mp b_{11}^*)$, $E_{3,4}$:$(0,\pm b_{22}^*,0, \pm b_{22}^*)$, where $b_{11}^*=\sqrt{\frac{\kappa+\omega^2-\omega_0^2}{3 \beta}}$ and $b_{22}^*=\sqrt{\frac{-\kappa+\omega^2-\omega_0^2}{3 \beta}}$.  The system also has asymmetric equilibrium points, which are $E_{5,6}$: $(0,\pm b^*_{33}, 0, \pm \frac{2 \kappa}{6\beta} b^{*^2}_{44} b^*_{33})$, $E_{7,8}$: $(0,\pm b^*_{44}, 0,\pm  \frac{2 \kappa}{6\beta}b^{*^2}_{33} b^*_{44})$, where $b^*_{33}$ $=\sqrt{\frac{(\omega^2-\omega_0^2)+\sqrt{-4 \kappa^2+ (\omega^2-\omega_0^2)^2}}{6 \beta}}$ and $,b^*_{44}$ $=\sqrt{\frac{(\omega^2-\omega_0^2)- \sqrt{-4 \kappa^2+ (\omega^2-\omega_0^2)^2}}{6 \beta}}$.  As $\omega=\omega_0- \Delta \omega$ and $\Delta \omega$ is a small deviation, $\omega^2-\omega_0^2$ is also small.  Thus, $b^*_{22}$, $b^*_{33}$ and $b_{44}^*$ cannot be real and the equilibrium points $E_{3,4}$, $E_{5,6}$ will not exist.  So, we confine our attention to the equilibrium points $E_{0}, E_1$ and $E_2$.

 Now, in order to investigate the stability of the above periodic solutions through a linear stability analysis, we obtain the eigenvalue equation as $A\chi_j=\lambda_j\chi_j$, where $\chi_j= [\xi_1\;\;\eta_1\;\;\xi_2\;\;\eta_2]^T$ and
\begin{eqnarray}
A=\left[ \begin{array}{cccc}
-\frac{3\beta a_1^* b_1^*}{ \omega}&c_{11}&0&-\frac{\kappa}{2\omega} \\
c_{12}&\frac{3\beta a^*_1 b^*_1}{ \omega}&\frac{\kappa}{2\omega}&0\\
0&-\frac{\kappa}{2\omega}&-\frac{3\beta a^*_2 b^*_2}{\omega}&c_{21}\\
\frac{\kappa}{2\omega}&0&c_{22}&\frac{3\beta a^*_2 b^*_2}{\omega}
\end{array} \right].
\end{eqnarray}
Here $c_{i1}$$=- \frac{3\beta ({a_i^*}^2+3{b_i^*}^2)}{2 \omega}+\frac{\omega^2-\omega_0^2}{2 \omega}$, $c_{i2}$$= \frac{3\beta (3{a_i^*}^2+{b_i^*}^2)}{2 \omega}-\frac{\omega^2-\omega_0^2}{2 \omega}$, $i=1,2$.   $\lambda_j$, $\chi_j$ ($j=1,2,3,4$) are the eigenvalues and eigenfunctions of the above eigenvalue equation.
 The eigenvalues of $A$ corresponding to the equilibrium point $E_0$: $(0,0,0,0)$ are
\begin{eqnarray}
\lambda_j=\pm i \frac{(-\kappa+(\omega^2 - \omega_0^2))}{2 \omega}, \pm i \frac{(\kappa+(\omega^2 - \omega_0^2))}{2 \omega}.
\end{eqnarray}
The eigenvalues of $A$ corresponding to $E_1$ and $E_2$ are
\begin{eqnarray}
\lambda_j=\pm \sqrt{\frac{-2 \kappa^2-\kappa(\omega^2-\omega_0^2)}{\omega}},0,0.
\end{eqnarray}
The eigenvalues of $A$ corresponding to $E_0$ are found to be neutrally stable always, whereas two of the eigenvalues associated with $E_1$ and $E_2$ are pure imaginary when $2 \kappa^2+\kappa (\omega^2-\omega_0^2)>0$.  When all the eigenvalues of $A$ corresponding to an equilibrium point are pure imaginary, the neutral stability of the equilibrium point will make the oscillation with frequency $\omega$ to be modulated by a slowly varying periodic amplitude.  It indicates that the system shows beats type oscillations.  As the equilibrium point $E_0$ is always neutrally stable, we have stable beats type periodic oscillations in the complete region $R_2$.  However, the equilibrium points $E_{1,2}$ have two of their eigenvalues as zero, and so one needs to include higher order corrections to conclusively decide about their stability.
\subsection{Case: $\Omega=\omega_0^2=0$}
In  this case, the existence of equilibrium points for different values of $\kappa$ is demonstrated in Table. \ref{tab}.  The eigenvalues of $J$ with respect to $e_0$ (Eq. (\ref{e0bip})) clearly show that it is always unstable.  The equilibrium points $e_{1}$ and $e_2$ are found to be neutrally stable for $\kappa>0$ and for the values of $\alpha$ specified in (\ref{c1bip}).  The equilibrium points $e_{3}$ or $e_4$ stabilize for $\kappa<0$ and give rise to oscillation death.

\subsection{Case $\Omega =\omega_0^2\leq 0$:}  
\begin{figure}
\begin{center}
\includegraphics[width=7 cm]{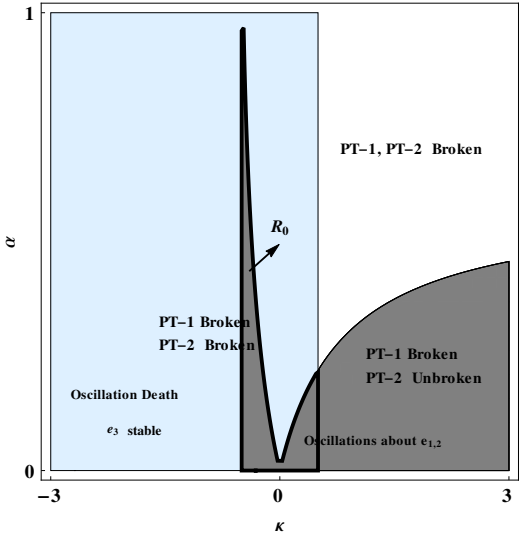}
\end{center}
\caption{(Color online) Phase diagram of (\ref{bip}) in ($\kappa, \alpha$) parametric space for $\Omega=\omega_0^2<0$. Figure is plotted for $\Omega=-1.0$, $\beta=1.0$, which shows the regions in oscillations about $e_1$ and $e_2$ exists (dark gray shaded region) and the region where oscillation death (light blue shaded region) occurs.  One can clearly note from the figure that the $\cal{PT}$$-1$ symmetry is broken everywhere. In the region denoted by $R_0$ (Region outlined by thick black line), we can find that there exists oscillations about $e_{1,2}$ and oscillation death occurs about $e_{3}$, thus $\cal{PT}$$-2$ symmetry is broken in the region. The gray shaded region excluding $R_0$ region gives rise to unbroken $\cal{PT}$$-2$ region.}
\label{odm1}
\end{figure}

Next, we wish to show the unbroken and broken $\cal{PT}$ regions corresponding to the system (\ref{bip}) with $\Omega=\omega_0^2<0$ or the double well potential case. The equilibrium points at different values of $\kappa$ for this case are also given in Table. \ref{tab}.  From the table, we can note that in contrast to the previous cases, in the region $-\omega_0^2 \leq \kappa \leq \omega_0^2$, the equilibrium points $e_{3,4}$ coexist with $e_{1,2}$.   From the results of the linear stability analysis of this case  (where $\Omega=\omega_0^2 <0$), we can find that the equilibrium point $e_0$ (see Eq. (\ref{e0bip}) Appendix \ref{apeg1}) completely loses its stability. Thus, when $\Omega<0$, as in the scalar case, $\cal{PT}$$-1$ symmetry is always broken.    The symmetric pair of equilibrium points  $e_{1}$, $e_{2}$ and  $e_3$, $e_4$  are still found to be stable in some regions in the $(\kappa, \alpha)$ parametric space. The region in which they are found to be neutrally stable or stable is given by Eqs. (\ref{c1bip}) and (\ref{e3bip}) and are shown by Fig.\ref{odm1}.  From the figure, we can observe that the $\cal{PT}$$-1$ symmetry is broken everywhere in the parametric space.  
\par Regarding the $\cal{PT}$$-2$ symmetry, Fig. \ref{odm1} shows the region in which the $\cal{PT}$$-2$ preserving fixed points $e_1$ and $e_2$ are neutrally stable (gray shaded region) and the region in which $\cal{PT}$$-2$ violating fixed point $e_3$ is stable (Light blue shaded regions).  All the regions in which $e_3$ is stable obviously correspond to the broken $\cal{PT}$$-2$ region.  Interestingly, in this case, there exists a region denoted by $R_0$ in Fig. \ref{odm1}, in which the stable region of $e_1$ and $e_2$ overlaps with the oscillation death region (stable region of the $\cal{PT}$$-2$ violating fixed point $e_3$).  Due to such coexistence, $\cal{PT}$$-2$ symmetry is broken in the region $R_0$. Thus the $\cal{PT}$$-2$ symmetry is unbroken only in the gray shaded region excluding $R_0$.

\section{\label{slin}Nonlinear plus linear damping}
\begin{figure*}[htb!]
\begin{center}
\vspace{-0.5cm}
\includegraphics[width=2\columnwidth]{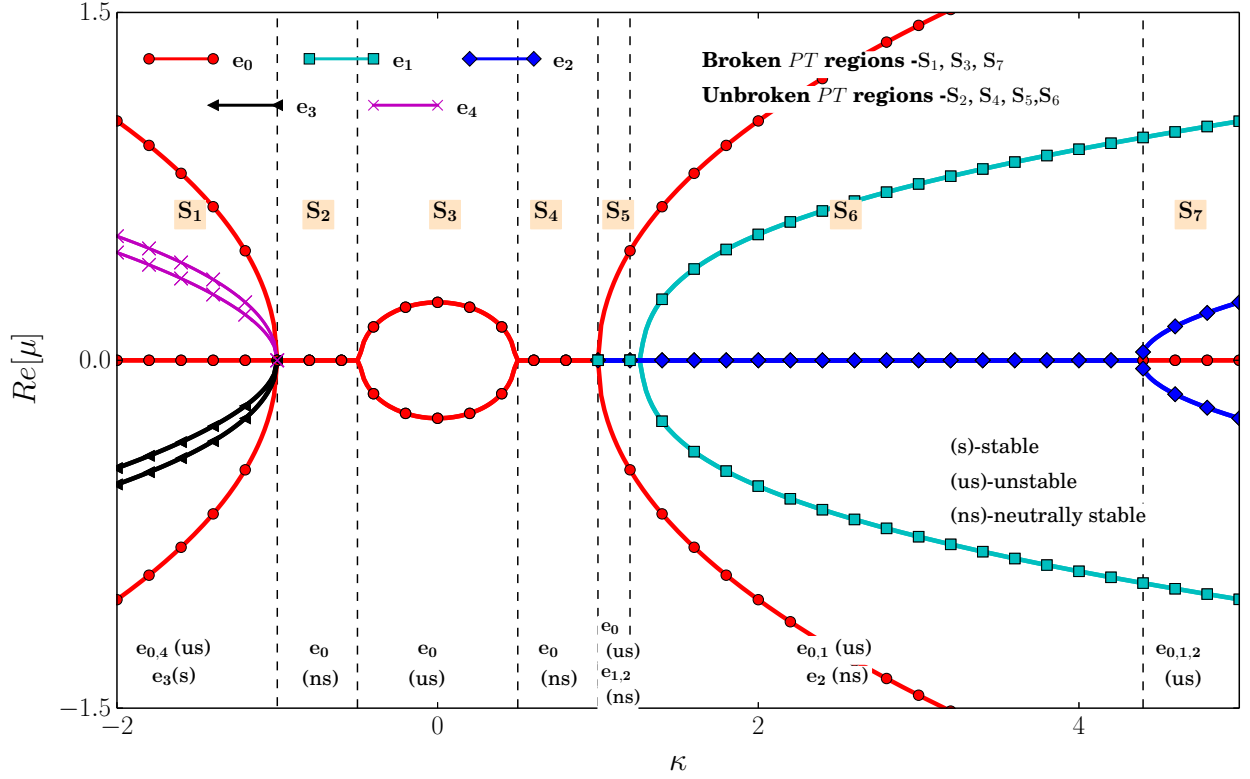}
\vspace{-0.5cm}
\end{center}
\caption{{(Color online) Linear stability of equilibrium points of (\ref{bip2}) for $\Omega=\omega_0^2>0$ given in Table. \ref{tab}.  Real parts of eigenvalues of $J$ given by Eq. (\ref{jac2}) are plotted as a function of $\kappa$ for the parameters $\gamma=0.5$, $\alpha=1.0$, $\beta=1.0$ and $\omega_0=1.0$.}} 
\label{je1}
\end{figure*}
\par Next we wish to investigate the effect of  the introduction of a linear damping on the dynamics of the nonlinearly damped system (\ref{bip}). For this purpose, let us introduce the linear damping terms in addition to the nonlinear damping introduced in Eq. (\ref{bip}).  Now, the system takes the form
\begin{eqnarray}
\ddot{x}+\gamma \dot{x}+\alpha x \dot{x}+\beta x^3+\omega_0^2 x+\kappa y=0, \nonumber \\
\ddot{y}-\gamma \dot{y}+\alpha y \dot{y}+\beta y^3+\omega_0^2 y+\kappa x=0,
\label{bip2}
\end{eqnarray}
where $\gamma$ is the linear loss/gain strength. Obviously, the added linear damping term in (\ref{bip2}) breaks the $\cal{PT}$$-1$ symmetry. Thus the system is only symmetric with respect to the $\cal{PT}$$-2$ operation.  Note that the equilibrium points of this system are the same as that of (\ref{bip}).  The stability determining Jacobian matrix in this case becomes 
\begin{small}
\begin{eqnarray}
J=\left[ \begin{array}{cccc}
0&1&0&0\\
c_{21}&-\gamma-\alpha x^*&-\kappa&0\\
0&0&0&1\\
-\kappa&0&c_{43}&\gamma-\alpha y^*\\
\end{array} \right],
\label{jac2}
\end{eqnarray}
\end{small}
where, $c_{21}=-\alpha x_1^*-3 \beta {x^*}^2 -\omega_0^2$, $c_{43}=-\alpha y_1^*-3 \beta {y^*}^2-\omega_0^2$. The eigenvalues of this Jacobian matix for different equilibrium points are given in Appendix \ref{apeg2}.  For simplicity, we take $\beta=1$ for further studies.  As in Sec. III, we look for $\cal{PT}$ regions of (\ref{bip2}) for the cases $\Omega=\omega_0^2>0$ and $\Omega =\omega_0^2\leq 0$ respectively.

\subsection{Case: $\Omega=\omega_0^2>0$}
\par To begin, we look for the $\cal{PT}$ regions of the system with respect to $\kappa$ for the case $\Omega=\omega_0^2>0$.  By fixing all the other parameters of the system as $\alpha=1.0$, $\gamma=0.5$, $\beta=1.0$ and $\omega_0^2=1.0$ in (\ref{bip2}), Fig. \ref{je1} shows the plot of the real part of eigenvalues of $J$ corresponding to the equilibrium points $e_0$, $e_{1}$, $e_{2}$, $e_{3}$ and  $e_{4}$ as $\kappa$ is varied.  It is divided into seven regions $S_1$, $S_2$, ..., $S_7$ along the $\kappa$-axis.  For the system (\ref{bip2}),  $\cal{PT}$$-2$ symmetry alone exists and the $\cal{PT}$ regions correspond to the regions in which the $\cal{PT}$$-2$ symmetry is unbroken.  The details are as follows.
\begin{figure}[htb!]
\begin{center}
\hspace{-0.2cm}
   \includegraphics[width=9.3cm]{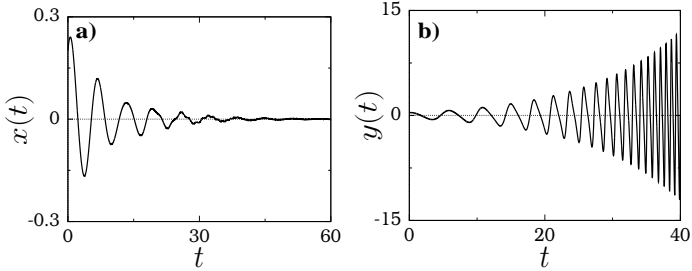}
\end{center}
   \caption{(Color online) Broken $\cal{PT}$ symmetry in the region $S_3$: Figures $(a)$ and $(b)$ are plotted for  $\kappa=0.01$, $\gamma=0.2$, $\alpha=1.0$, $\omega_0$$=1.0$ and $\beta=1.0$ that show the time series plots of $x$ and $y$.  The damped and growing oscillations of $x(t)$ and $y(t)$ indicate that for finite values of $\kappa$, the $\cal{PT}$$-2$ symmetry is broken.}
\label{jf4}   
\end{figure}
\begin{itemize}
\item In the region $S_1$ of Fig. \ref{je1}, where $\kappa<-\omega_0^2=-1.0$, we can find that among the equilibrium points $e_0$, $e_3$ and $e_4$, only $e_3$ is found to be stable which leads to oscillation death.  As mentioned in the previous case, the $\cal{PT}$$-2$ symmetry is broken in this region. 
\item The region corresponding to the values of $\kappa$ between $-\omega_0^2<\kappa< \omega_0^2$ ($-1<\kappa< 1$) is now divided into three regions, namely  $S_2$, $S_3$ and $S_4$.  In these regions as mentioned in Table. \ref{tab}, the equilibrium point $e_0$ alone exists.
\begin{enumerate}[(a)]
\item In the region $S_2$, where $-\omega_0^2 < \kappa \leq -\sqrt{\frac{4 \omega_0^4-(2\omega_0^2 - \gamma^2)^2}{4}}$ (that is $-1 \leq \kappa \leq -0.484$), we can note that the equilibrium point $e_0$ is found to be neutrally stable (which can also be seen from Eq. (\ref{kapb2})) and gives rise to an unbroken $\cal{PT}$ region.

\item In the region $S_3$, where $\kappa$ takes smaller values, $-\sqrt{\frac{4 \omega_0^4-(2\omega_0^2 - \gamma^2)^2}{4}} \leq \kappa \leq \sqrt{\frac{4 \omega_0^4-(2\omega_0^2 - \gamma^2)^2}{4}}$ (that is $-0.484 \leq \kappa \leq 0.484$), we can see that the equilibrium point $e_0$ loses its stability (can be seen also from Eq. (\ref{kapb2})) and the $\cal{PT}$$-2$ symmetry is broken now.  As this $\cal{PT}$$-2$ symmetry appears because of coupling (that is, the $\cal{PT}$$-2$ symmetry disappears when $\kappa=0$) it will not be preserved for smaller values of $\kappa$.  Figs. \ref{jf4}(a) and \ref{jf4}(b) are plotted in the region, which shows the damped oscillation in $x$ and grow up oscillation $y$ which shows the unbalanced energy between the $x$ and $y$ oscillators. 

\item Now increasing $\kappa$, in the region $S_4$, for $\sqrt{\frac{4 \omega_0^4-(2\omega_0^2 - \gamma^2)^2}{4}} \leq \kappa < \omega_0^2$, $e_0$ again becomes neutrally stable and gives rise to unbroken $\cal{PT}$ region.
\end{enumerate}
\item For $\kappa>\omega_0^2=1$, there exists three regions which are designated as $S_5$, $S_6$ and $S_7$, identified from Eq. (\ref{e1b2}). In these regions, the equilibrium points $e_0$, $e_1$ and $e_2$ are found to exist (see Table-\ref{tab}).

\begin{enumerate}[(a)]
\item In the region $S_5$, ($1.0 < \kappa <1.28$), the equilibrium point $e_0$ is found to be unstable, but $e_1$ and $e_2$ are found to be neutrally stable (can be seen also from Eq. (\ref{e1b2})).  As these equilibrium points traces itself upon $\cal{PT}$$-2$ operation (that is $\cal{PT}$$-2$[$e_1$]=$e_1$), the $\cal{PT}$$-2$ symmetry in the region is said to be unbroken.  

\item In the region $S_6$, ($1.28< \kappa< 4.4$), in addition to $e_0$, $e_1$ also loses its stability (can be seen also from Eq. (\ref{e1b2})).   But $e_2$ is still neutrally stable, thus the region again corresponds to an unbroken $\cal{PT}$ region.\\

\item On further increasing $\kappa$, for $\kappa> 4.4$, in the region $S_7$, all the equilibrium points $e_0$, $e_1$ and $e_2$ become unstable.  Thus $\cal{PT}$ is broken for higher values of $\kappa$.
\end{enumerate}
\end{itemize}
 
\begin{figure}
\begin{center}
\includegraphics[width=0.85\linewidth]{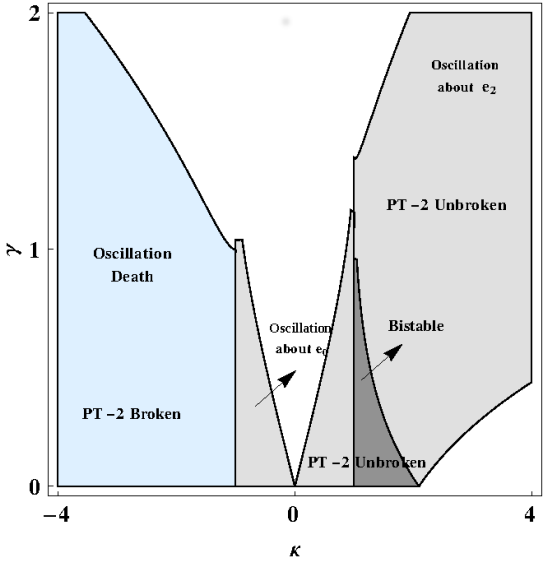}
\end{center}
\caption{(Color online) Broken and unbroken $\cal{PT}$ regions corresponding to the system (\ref{bip2}) in the ($\kappa, \gamma$) parametric space for $\Omega=\omega_0^2>0$, which is plotted for $\alpha=1.0$, $\omega_0=\beta=1.0$.  The light blue shaded region corresponds to the oscillation death region. The light and dark gray shaded regions denote the unbroken $\cal{PT}$$-2$ region.  The dark gray shaded region corresponds to the bistable region in the sense that the equilibrium points $e_1$ and $e_2$ are neutrally stable in the region.}
\label{inh2}
\end{figure}

\par For $\alpha=1.0$, $\omega_0=1.0$, and $\beta=1.0$, the broken and unbroken regions in the ($\kappa,\gamma$) parametric space of (\ref{bip2}) are indicated in Fig. \ref{inh2}.  By comparing Fig. \ref{inh2} with Fig. \ref{alp}, we can find the appearance of oscillation death for the values $\kappa<-1$ as in the previous case (\ref{bip}).  But in contrast to the previous case, the oscillation death regime disappears with an increase of $\gamma$.   By increasing $\kappa$, the unbroken $\cal{PT}$ region appears in the range $-\omega_0^2 \leq \kappa \leq \omega_0^2 $ (where $\omega_0=1.0$).  In this region by increasing $\gamma$, the system shows symmetry breaking (see Eqs. (\ref{e0b2}) in Appendix \ref{apeg2}).  But in the previous case (\ref{bip}), we cannot find this type of behavior, where the $\cal{PT}$ symmetry is never broken by increasing the loss/gain strength $\alpha$ (see Fig. \ref{alp} and Eq. (\ref{e0bip})).  
\par For $\kappa>1$ (the region in which $e_1$ and $e_2$ appear), Fig. \ref{inh2} indicates that when $\kappa$ is smaller than $\approx 2.2$, the $\cal{PT}$ symmetry of the system is preserved for lower values of $\gamma$ and it is broken for higher values of $\gamma$. Increasing $\kappa$ beyond $\approx 2.2$, the $\cal{PT}$ symmetry of the system is broken for lower values of $\gamma$, and on increasing $\gamma$ the $\cal{PT}$ symmetry is restored or it becomes unbroken for the values of $\gamma$ mentioned in Eq. (\ref{c2b2}).  On further increasing $\gamma$, the symmetry is again broken. Generally, in the standard type of $\cal{PT}$-symmetric systems, $\cal{PT}$ is unbroken for lower values of $\gamma$ and broken for higher values of $\gamma$. Thus, this type of $\cal{PT}$ restoration with the increase of loss/gain strength is unusual compared to the general $\cal{PT}$-symmetric systems,   except for the case of Aubry- Andre model with two lattice potentials \cite{k3,aub}.  As mentioned in the introduction, the latter model is a lattice model in which the lattice potential is applied in such a way that each element of the lattice has different amount of loss and gain that makes the loss and gain present in the lattice to be  position dependent.  Then, the  phenomenon of $\cal{PT}$ restoration at higher values of loss/gain strength appears only when two such lattice potentials are applied simultaneously. The reason for this  type of $\cal{PT}$ restoration is the competition between the two applied potentials which introduces loss and gain in the system \cite{k3}. 
\par Similarly, in our case if a single damping is present in the system (\ref{bip}), we cannot observe such $\cal{PT}$ restoration at higher loss/gain strength (see Fig. \ref{alp}).   But when two or more types of damping present in the system, as in the case of (\ref{bip2}) (where  linear and nonlinear dampings are present in the system) we can observe this type of $\cal{PT}$ restoration (see Fig. \ref{inh2}).  The above point will be further discussed in detail in the next section, where we will also show that by properly choosing the form of nonlinear damping, we can also tailor the $\cal{PT}$ regions of the system in the parametric space.  Fig. \ref{inh2} shows that there exists bistable regions for finite values of $\gamma$ and by increasing the coupling strength $\kappa$ the bistable region disappears. 
\subsection{Rotating wave approximation}
\par Now, we look for the stability of the periodic orbits about $e_0$ in the region $-\omega_0^2\leq \kappa \leq \omega_0^2$. As we did in the previous case (\ref{bip}), we find that the amplitude equations are 
\begin{small}
\begin{eqnarray}
\dot{R}_1&=&\frac{1}{2i{\omega} }(-i \gamma \omega R_1-3\beta|R_1|^2 R_1+(\omega^2-\omega_0^2)R_1-\kappa R_2), \;\;\; \nonumber \\
\dot{R}_2&=&\frac{1}{2i{\omega} }(i \gamma \omega R_2-3\beta|R_2|^2 R_2+(\omega^2-\omega_0^2)R_2-\kappa R_1). \;\;\;
\label{phi2}
\end{eqnarray}
\end{small}
Now, separating the real and imaginary parts of the equation as $R_1=a_1+ib_1$, $R_2=a_2+ib_2$, and from the linear stability analysis of the above equation, we can find that the system has an equilibrium point ($0,0,0,0$), whose
eigenvalues are
\begin{small}
\begin{eqnarray}
\lambda=\pm\frac{1}{2 \omega}\sqrt{ \left(-(\kappa^2-\gamma^2\omega^2)-(\omega^2-\omega_0^2)^2 
\pm 2 \sqrt{c_1} \,\right)} 
\end{eqnarray}
\end{small}
where $c_1=(\kappa^2-\gamma^2\omega^2)(\omega^2-\omega_0^2)^2$.  The equilibrium points are found to be neutrally stable for $-\sqrt{\frac{\kappa}{\omega}}$ $\leq $ $\gamma$ $\leq$ $\sqrt{\frac{\kappa}{\omega}}$. The linear stability discussed in the previous section tells that the equilibrium point $e_0$ can become neutrally stable in the region given by Eq. (\ref{add}) (see Appendix \ref{apeg2}) and the above stability analysis of periodic orbits in the region shows that the oscillations are found to be stable only for the values of $\gamma$ mentioned above.

\subsection{Case: $\Omega=\omega_0^2\leq 0$}
\par
\begin{figure}
\begin{center}
\includegraphics[width=7.5cm]{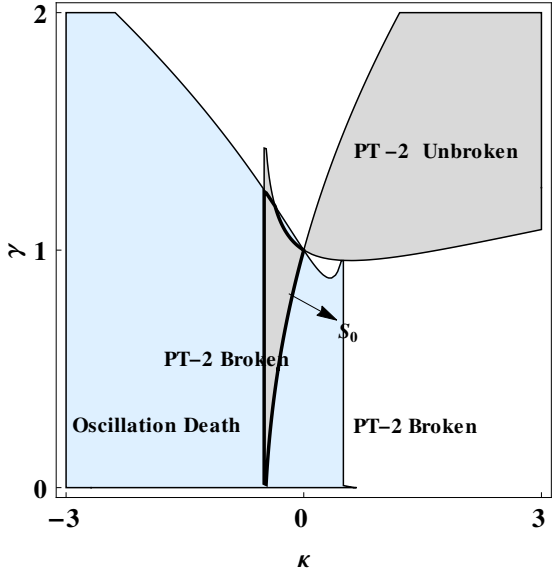}
\end{center}
\caption{(Color online) Broken and unbroken $\cal{PT}$ regions corresponding to the system (\ref{bip2}) in the ($\kappa, \gamma$) parametric space for $\Omega=\omega_0^2<0$, which is plotted for $\alpha=1.0$, $\Omega=-1.0$, $\beta=1.0$.  The light blue and light gray shaded regions correspond to oscillation death (stable region of $e_3$) and neutrally stable region of $e_2$ respectively.  In the region $S_0$, the region of stable region of $e_2$ coexists with the stable region of $e_3$, thus $\cal{PT}$ symmetry is broken in the region.  The unbroken region corresponds to the gray shaded region excluding $S_0$.}
\label{odm}
\end{figure}

 By taking $\Omega=\omega_0^2\leq 0$, the equilibrium point $e_0$ loses its stability (see Eq. \ref{e0b2}).  The equilibrium points $e_{1,2}$ and $e_{3,4}$ alone are found to be stable and the stable regions of these equilibrium points are given in Appendix \ref{apeg2}.  Similar to the previous case, we have observed a region denoted by $S_0$ in Fig. \ref{odm}, in which a neutrally stable $\cal{PT}$ preserving fixed point ($e_2$) coexists with $\cal{PT}$ violating fixed points. Thus this region $S_0$ corresponds to broken $\cal{PT}$ region. The gray shaded region excluding $S_0$ alone corresponds to the unbroken $\cal{PT}$ region.  As in the case where $\Omega>0$, here also $\cal{PT}$ restoration at higher loss/gain occurs.

\section{\label{gene}General Case }
\par In this section, we consider a more general coupled $\cal{PT}$-symmetric cubic anharmonic oscillator system with nonlinear damping.  Here, we take the nonlinear damping term $h(x,\dot{x})$ to be of the form $f(x)\dot{x}$ so that the equation of motion will take the form
\begin{eqnarray}
&&\ddot{x}+\gamma \dot{x}+(-1)^n \alpha f(x)\dot{ x}+\beta x^3+\omega_0^2  x+\kappa y=0,  \nonumber \\
&&\ddot{y}-\gamma \dot{y}+ \alpha f(y)\dot{y} +\beta y^3 +\omega_0^2 y+\kappa x=0, 
\label{bip3}
\end{eqnarray}
where $n=0$ if $f(x)$ is an odd function and $n=1$ if $f(x)$ is  even.  Thus the system is $\cal{PT}$-symmetric with respect to the $\cal{PT}$$-2$ operation.  The novel bi-$\cal{PT}$-symmetric case arises when $f(x)$ is odd and $\gamma=0$.  For all forms of $f(x)$, the equilibrium points are found to be the same as that of (\ref{bip}).  Now through the linear stability analysis let us find the unbroken and broken $\cal{PT}$-symmetric regions.  The Jacobian matrix corresponding to (\ref{bip3}) is 
\begin{small}
\begin{eqnarray}
J=\left[ \begin{array}{cccc}
0&1&0&0\\
c_{21}&-\gamma-(-1)^n \alpha f(x^*)&-\kappa&0\\
0&0&0&1\\
-\kappa&0&c_{43}&\gamma- \alpha f(y^*)\\
\end{array} \right],
\label{jac3}
\end{eqnarray}
\end{small}
where $c_{21}=-\alpha f'(x^*) x_1^*-3 \beta {x^*}^2 -\omega_0^2$, $c_{43}=-\alpha f'(y^*) y_1^*-3 \beta {y^*}^2-\omega_0^2$.  For simplicity, we consider the case of $\omega_0=1$, $\beta=1$. The eigenvalues of this Jacobian matrix corresponding to odd and even $f(x)$ cases of the system (\ref{bip3}) about various equilibrium points are given in Appendix \ref{new}. 
\begin{figure*}[htb!]
\begin{center}
\hspace{-0.5cm}
   \includegraphics[width=1.5\columnwidth]{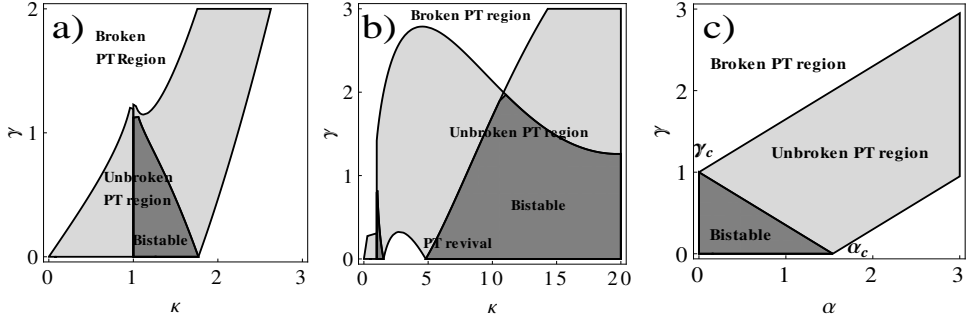}
\end{center}
   \caption{(Color online) Phase diagram in ($\kappa, \gamma$) space:  $(a)$ and $(b)$ denote broken (white), unbroken (gray) and bistable (dark gray) regions with $f(x)=x^3$ and $f(x)=sin x$, respectively, for $\alpha=1.5$.  (c): Phase diagram in  ($\alpha, \gamma$) space corresponding to $f(x)=sinx$ and $\kappa=1.5$}
\label{odd}   
\end{figure*}
\subsection{Case: $f(x)$ is odd}
   Considering the case where $f(x)$ is an odd function, in the region $-1 \leq \kappa \leq 1$ (see Table \ref{tab}), in which the equilibrium point $e_0$ alone exists, the corresponding eigenvalues of $J$ are the same as in (\ref{e0b2}).  In this region, we can find that the eigenvalues do not depend on $\alpha$ but depends on $\gamma$ (see Eq. (\ref{e0b2})).  The region of unbroken $\cal{PT}$ symmetry is  confined to 
\begin{eqnarray}
-\sqrt{2-2\sqrt{1-\kappa^2}} \leq  \gamma \leq \sqrt{2-2\sqrt{1-\kappa^2}}.
\label{e0}
\end{eqnarray}
From the above, it is clear that when $\gamma=0$ the $\cal{PT}$ is always unbroken for all the values of $\alpha$ in the region $-1\leq \kappa \leq 1$.  This indicates that in a purely nonlinearly damped system, we cannot observe any symmetry breaking while varying the nonlinear damping strength ($\alpha$) in this region.  By varying $\gamma$, we observe symmetry breaking for higher values of $|\gamma|>\sqrt{2-2\sqrt{1-\kappa^2}}$.
\par  In the region $\kappa>1$, where the non-trivial equilibrium points $e_1$ and $e_2$ come into action, we will show that by properly choosing the nonlinear damping we can tailor the $\cal{PT}$ regions.  In this regime, for the case in which $f(x)$ is an odd function, the unbroken $\cal{PT}$ region lies within the range of $\gamma$ specified by (see Eq. (\ref{c2bg}) in Appendix \ref{new})
\begin{eqnarray}
\pm \alpha f(\sqrt{\kappa-1})-\sqrt{a_1} \leq \gamma \leq \pm \alpha f(\sqrt{\kappa-1})+\sqrt{a_1},
\label{g1}
\end{eqnarray}
where $a_1=(6\kappa-4)-4\sqrt{(2\kappa-1)(\kappa-1)}$.  The presence of the term $\alpha f(\sqrt{\kappa-1})$ in the above equation is found to be important.  Because considering the case where $\alpha f(\sqrt{\kappa-1}) = 0$, the $\cal{PT}$ symmetry is unbroken for lower values of $\gamma$ specified by $|\gamma|< \sqrt{a_1}$ and is broken for the higher values of $\gamma$ specified by $|\gamma|> \sqrt{a_1}$. But, in the case where $\alpha f(\sqrt{\kappa-1}) \neq 0$, for the values of $\gamma$ defined by $0< |\gamma| < \alpha f(\sqrt{\kappa-1})-\sqrt{a_1}$, the $\cal{PT}$ symmetry is broken while it is unbroken for the values of $\gamma$ defined by (\ref{g1}).  Thus, here the $\cal{PT}$ symmetry breaking occurs at lower values of $\gamma$ and the restoration of symmetry occurs by increasing $\gamma$.  We can also note that the term $\alpha f(\sqrt{\kappa-1})$ depends on the form of $f(x)$, which helps in tailoring $\cal{PT}$ regions of the system.

\par In Fig. \ref{odd}, we have presented the $\cal{PT}$ regions of the system for the cases $f(x)=x^3$ and $f(x)=sin x$, which clearly show that the $\cal{PT}$ regions can be tailored with the systems of the type (\ref{bip3}) by properly choosing the form of $f(x)$.  From Fig. \ref{odd}(b), we can note that by choosing $f(x)$ to be a periodic one, we can observe $\cal{PT}$ revivals.  
\par In Fig. \ref{odd}(c), we have shown the $\cal{PT}$ regions of the system in the ($\gamma, \alpha$) parametric space corresponding to the $f(x)=sin x$ case, while the figure looks qualitatively the same for $f(x)=x^3$. The figure indicates that increasing $\gamma$ (or $\alpha$) beyond a critical value, denoted as $\gamma_c$ (or $\alpha_c$), the unbroken $\cal{PT}$ region appears only when $\alpha$ (or $\gamma$) is also sufficiently large.  
\subsection{Case: $f(x)$ is even}
 The case of even $f(x)$ can again be divided into two sub-cases: (i) $f(0)=0$ and (ii) $f(0)=$ a nonzero constant say, $1$ (For the odd $f(x)$ case, $f(0)=0$ always and so there are no sub-cases.) \\
\begin{figure*}[ht]
\begin{center}
   \includegraphics[width=1.5\columnwidth]{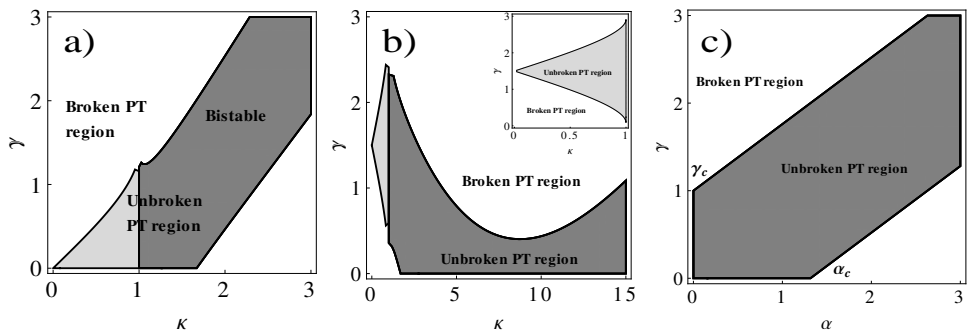}
\end{center}
   \caption{(Color online) Phase diagram in ($\kappa, \gamma$) space: $(a)$ and $(b)$ denote broken (white), unbroken (gray) and bistable (dark gray) regions with $f(x)=x^2$ and $f(x)=cos x$ (where $\alpha=1.5$).  The inset in  Fig. (b) shows the $\cal{PT}$ regions corresponding to the case $f(x)=cos x$ for values of $\kappa$ between $0<\kappa<1$ in ($\kappa, \gamma$) space.  (c): Phase diagram in  ($\alpha, \gamma$) space corresponding to $f(x)=cos x$ and $\kappa=1.5$}
\label{aal}
 \end{figure*}
\par {\bf Case (i) $f(0)=0$:}~~~ Considering the case of $f(x)$ with $f(0)=0$ (Example: $f(x)=x^2$), in the region $-1 \leq \kappa \leq 1$ where the equilibrium point $e_0$ alone exists (see Table-\ref{tab}), the corresponding eigenvalues of $J$ (given in (\ref{jac3})) are found to be the same as in (\ref{e0b2}) and the unbroken $\cal{PT}$ regions of the system are also the same as that of (\ref{e0}).\\

\par {\bf Case (ii) $f(0)=1$:}~~~In this case, for example $f(x)=cos x$ or $e^{-x^2}$, the eigenvalues of $J$ are different from case (i) and they are given in (\ref{e0bg2}).  In contrast to the previous cases, the eigenvalues of $J$ corresponding to $e_0$ are found to depend on $\alpha$, see Eq. (\ref{e0bg2}), and the $\cal{PT}$ unbroken region can be given in terms of $\gamma$ as
\begin{eqnarray}
\alpha-\sqrt{a_2}\leq\gamma\leq\alpha+\sqrt{a_2}.
\label{cos}
\end{eqnarray}
where $a_2=2\omega_0^2-\sqrt{4(1-\kappa^2)}$.
This equation indicates that the $\cal{PT}$ symmetry is found to be broken for values of $\gamma$ outside the range specified by (\ref{cos}) and $\cal{PT}$ symmetry becomes unbroken by choosing $\gamma$ within the range given in (\ref{cos}).  Thus the $\cal{PT}$ symmetry is broken for lower values of $\gamma$, $\gamma<\alpha-\sqrt{a_2}$ and restored at higher $\gamma$, as in Eq. (\ref{cos}). 
\par  As $\cal{PT}$ is broken for $\gamma< \alpha-\sqrt{a_2}$, for $\alpha>0$ the $\cal{PT}$ regions preferentially exist for $\gamma>0$ and found to be scarce for $\gamma<0$.  In other words, the unbroken $\cal{PT}$ regions are abundant, if the loss due to the linear (or nonlinear) damping is introduced in the $x$-oscillator and the loss due to the nonlinear (or linear) damping is introduced in the $y$- oscillator.  When loss (or also gain) due to both the linear and nonlinear damping is introduced in the same oscillator, the unbroken $\cal{PT}$ regions become scarce.
\par Now considering the region ($\kappa>1$), where the non-trivial equilibrium points exist (see Table-\ref{tab}), the dynamics corresponding to the two sub-cases (case (i) and case (ii)) are the same.  The eigenvalues of $e_1$ and $e_2$ are given in (\ref{e1bg2}), which become purely imaginary in the region
\begin{eqnarray}
 \alpha f(\sqrt{\kappa-1})-\sqrt{a_1} \leq \gamma \leq \alpha f(\sqrt{\kappa-1})+\sqrt{a_1},
\label{g2}
\end{eqnarray}
where $a_1=(6\kappa-4)-4\sqrt{(2\kappa-1)(\kappa-1)}$.  Comparing the above with the one corresponding to the $f(x)$ odd case (see Eqs. (\ref{g1}) and (\ref{g2})), we can find that in this case the unbroken $\cal{PT}$ regions are scarce for $\gamma<0$.  The presence of the term $\alpha f(\sqrt{\kappa-1})$ indicates that the $\cal{PT}$ restoration can occur at higher values of loss/gain, which confirms that the $\cal{PT}$ regions can be tailored by a proper choice of $f(x)$. 

Fig. \ref{aal}(a) shows the $\cal{PT}$ regions of the system (\ref{bip3}) for the choice of $f(x)=x^2$ which corresponds to the sub-case (i) $f(0)=0$.  Fig. \ref{aal}(b) is plotted for $f(x)=cos x$, corresponding to the sub-case (ii), namely, $f(0)=1$.  The inset in the figure clearly shows that in this system even for $\kappa<1$ the $\cal{PT}$ restoration at higher loss/gain strength occurs.  Figs. \ref{aal}(a) and \ref{aal}(b) clearly show that the $\cal{PT}$ regions can be tailored by the proper choice of $f(x)$. Fig. \ref{aal}(c) show the $\cal{PT}$ regions in the ($\gamma, \alpha$) parametric space for the choice $f(x)=\cos x$, which shows the existence of critical values $\gamma_c$ and $\alpha_c$ above which the $\cal{PT}$ is unbroken for higher loss/gain strength.
\section{\label{conc}conclusion}
\par In this work, we have brought out the nature of the novel bi-$\cal{PT}$ symmetry of certain nonlinear systems with position dependent loss-gain profiles.  We have pointed out that the $\cal{PT}$-symmetric cases of this type of nonlinear systems with position dependent loss-gain profile occur even with a single degree of freedom.  These scalar nonlinear $\cal{PT}$-symmetric systems are also found to show $\cal{PT}$ symmetry breaking.  We have demonstrated the nature of $\cal{PT}$-symmetry preservation and breaking with an interesting integrable example of damped nonlinear system.  By coupling two such scalar $\cal{PT}$-symmetric systems in a proper way, we have shown the existence of the novel bi-$\cal{PT}$-symmetric systems in two dimensions.  We have also illustrated the phenomenon of symmetry breaking of the two $\cal{PT}$ symmetries in this bi-$\cal{PT}$-symmetric system.  When this system is acted upon by a single nonlinear damping, we observed that for smaller coupling strengths, the coupled system shows no symmetry breaking while varying nonlinear loss/gain strength, whereas the coupled $\cal{PT}$-symmetric system with a linear damping  \cite{b8} shows symmetry breaking by increasing loss/gain strength.  By strengthening the coupling, this nonlinearly damped system shows symmetry breaking for higher loss/gain strength.  Then, by applying the linear damping in addition to the nonlinear damping in a competing way, our results show that as in the $\cal{PT}$-symmetric Aubry-Andre model, $\cal{PT}$ restoration at higher values of loss/gain strength occurs.  The advantage of having position dependent nonlinear damping with a competing linear damping is to help to tailor the $\cal{PT}$ regions of the system according to the needs by properly designing the nonlinear loss and gain profile.  We have also observed $\cal{PT}$ revivals in the systems which have loss and gain periodically in space.

\section*{Acknowledgement}
SK thanks the Department of Science and Technology (DST), Government of India, for providing a INSPIRE Fellowship. The work of VKC forms part of a research project sponsored by INSA Young Scientist Project.  The work of MS forms part of a research project sponsored by Department of Science and Technology, Government of India.  The work forms part of an IRHPA project of ML, sponsored by the Department of Science Technology (DST), Government of India, who is also supported by a DAE Raja Ramanna Fellowship. 
\appendix 
 
\section{\label{appe1}Symmetry breaking in a $\cal{P}$- symmetric cubic anharmonic oscillator}
Here, we demonstrate the $\cal{P}$-symmetry breaking in a cubic anharmonic oscillator through the solution of its IVP.  Let us consider the cubic oscillator equation
\begin{eqnarray}
\ddot{x}+\lambda x+\beta x^3=0, \quad \lambda=\omega_0^2.
\label{duffy}
\end{eqnarray}
For simplicity, we consider $\beta>0$ for further discussions.  The $\cal{P}$ symmetry breaking in such a system is well known in the literature.  For $\lambda>0$, this system has an equilibrium point $e_0$:$(0,0)$ and the equilibrium point is found to be neutrally stable. As $\cal{P}$[$(0,0)$]=$(0,0)$, the $\cal{P}$ symmetry in the region is unbroken. By decreasing $\lambda$ to $\lambda<0$, the equilibrium point $e_0$ loses its stability and gives birth to two new neutrally stable equilibrium points which are $e_{1,2}$:$(\pm \sqrt{\frac{-\lambda}{\beta}},0)$.  In fact $e_0$ is a saddle and $e_{1,2}$ are centre type equilibrium points.  But these new equilibrium points $e_1$ and $e_2$ do not preserve symmetry as $\cal{P}$$(e_1)=e_2$ and vice versa.  Thus $\cal{P}$ symmetry is broken while $\lambda<0$.  All the stable equilibrium points correspond to minimum energy values. 
\par The system is an integrable one and its exact solution is also available in the literature \cite{new11,new12}.  Now, we demonstrate the above $\cal{P}$ symmetry breaking from the solution of the IVP of the system. 
\par Here the general solution of the system is given as follows \\

\underline {Case-$1$: $\lambda>0$:}
\begin{eqnarray}
x(t)=A cn[\Omega t+\delta,k]
\label{lamg0}
\end{eqnarray}
where $\Omega=\sqrt{\omega_0^2+\beta A^2}$, the square of the modulus $k^2=\frac{\beta A^2}{2 (\omega_0^2+\beta A^2)}$ and $\delta$ is a constant. The associated energy integral is 
$E=H=\frac{1}{2} \dot{x}^2+\frac{1}{2} \omega_0^2 x^2+\frac{1}{4} \beta x^4$ $=\frac{1}{2} \omega_0^2 A^2+\frac{1}{4} \beta A^4$.  Then considering without loss of generality the IVP, $x(0)=A$, $\dot{x}(0)=0$, in order that $\cal{P}$$x(0)$ $=$ $x(0)$, $\cal{P}$$\dot{x}(0)=\dot{x}(0)$ $\Rightarrow$ $A=-A$ which is possible only if $A=0$.  Further since one requires $\cal{P}$$[x(t)]$ $=-x(t)$ $\Rightarrow$ $x(t)=-x(t)$.  From (\ref{lamg0}), only the possibility $A=0$ $\Rightarrow$ $x(t)=0$, $\dot{x}(t)=0$ for all $t \geq 0$ is the admissible solution of the IVP which preserves $\cal{P}$ symmetry.  The corresponding energy $E=0$ has the minimum value.  The excited states of the system may said to be $\cal{P}$ symmetric if the time translation is included.\\

\underline{Case-$2$: $\lambda<0$:}
\par On the other hand one finds the following general solutions for the case $\lambda<0$ in Eq. (\ref{duffy})

(i) $0 \leq A \leq \sqrt{\frac{|\lambda|}{\beta}}$ :
\par In this range only the trivial solution exists. 
\begin{eqnarray}
 x(t)=0, \dot{x}=0
\label{eqst}
\end{eqnarray}
(ii) $\sqrt{\frac{|\lambda|}{\beta}} \leq A \leq \sqrt{\frac{2 |\lambda|}{\beta}}$: 
\par In this region we have the following two distinct periodic solutions in the two wells
\begin{eqnarray}
x(t)&=&\pm A dn(\Omega t+\delta,k) \\
\dot{x}(t)&=&\mp A \Omega k^2 sn(\Omega t+\delta,k) cn (\Omega t+\delta, k)
\label{dn}
\end{eqnarray}
where $\Omega^2=\frac{\beta A^2}{2}$ and $k^2=\frac{2(\beta A^2-|\lambda|)}{\beta A^2}$, and $\delta$ is a constant.\\
(iii) $A \geq \sqrt{\frac{2 |\lambda|}{\beta}}$:
\par In this region, one has the solution
\begin{eqnarray}
x(t)&=&A cn(\Omega t+\delta), \label{cn1} \\
\dot{x}(t)&=&-A \Omega sn (\Omega t+\delta,k) dn(\Omega t+\delta,k), \label{sndn2} \\
\Omega&=& \sqrt{-|\lambda|+\beta A^2},   \qquad k^2=  \frac{\beta A^2}{2(-|\lambda|+\beta A^2)}.     \nonumber 
\label{cnm}
\end{eqnarray}
Considering the IVP $x(0)=A$, $\dot{x}(0)=0$, one again finds $x(t)=0$, $\dot{x}(t)=0$ is the only possible $\cal{P}$-symmetric solution, existing when $A<\sqrt{\frac{|\lambda|}{\beta}}$.  But in the region $\sqrt{\frac{|\lambda|}{\beta}}< A< \sqrt{\frac{2 |\lambda|}{\beta}}$,  one also has the non-trivial distinct set of solutions
\begin{eqnarray}
x_1(t)&=&+A dn(\Omega t, k), \label{dn1} \\
x_2(t)&=&-A dn(\Omega t, k),  \label{dn2}
\end{eqnarray}
such that
\begin{eqnarray}
{\cal{P}}x_1(t)=x_2(t) \;\mathrm{and} \; {\cal{P}}x_2(t)=x_1(t),
\label{x1x2}
\end{eqnarray} 
and so also
\begin{eqnarray}
{\cal{P}}\dot{x}_1(t)=\dot{x}_2(t) \;\mathrm{and} \; {\cal{P}}\dot{x}_2(t)=\dot{x}_1(t).
\label{x1x22}
\end{eqnarray} 
Note that the value of the corresponding energy integral $E=-\frac{1}{2} |\lambda|   A^2+\frac{1}{4} \beta A^4$ and its minimum value $E=E_{min}=-\frac{1}{4} \frac{|\lambda|^2}{\beta}$ is attained when the amplitude $A=\sqrt{\frac{|\lambda|}{\beta}}$.  In this case, the square of the modulus $k=0$, $\Omega=0$ and so $dn(u,0)=1$ and 
\begin{eqnarray}
x(t)=\pm \sqrt{\frac{|\lambda|}{\beta}}, \dot{x}(0)=0.
\label{eqptl}
\end{eqnarray}
Note that the solution (\ref{dn1}, \ref{dn2}), including the limiting case, all correspond to energies lower than the $\cal{P}$-symmetric state $E_0=(0,0)$ and break the $\cal{P}$ symmetry.
\par
\begin{figure}[htb!]
\begin{center}
   \includegraphics[width=9.0cm]{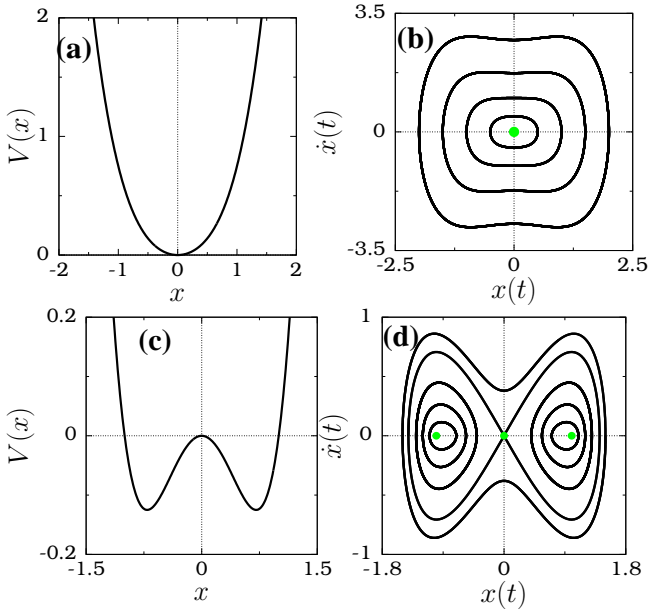}
\end{center}
   \caption{(Color online) (a) and (b) Single well: Potential energy curve and phase portrait of the system (\ref{duffy}) for $\lambda=1$ and $\beta=1$.  (c) and (d) Double well: potential energy and phase portrait of the system (\ref{duffy}) for $\lambda=-1$ and $\beta=1$. }
\label{duffg}
 \end{figure}

 Finally in the region $A \geq \sqrt{\frac{2|\lambda|}{\beta}}$, there exists no $\cal{P}$-symmetric solution, unless time translation and time reversal symmetries are also allowed in which the cases the phase trajectories are closed with $A \geq \sqrt{\frac{2 |\lambda|}{\beta}}$.  The associated phase trajectories are presented in Fig. \ref{duffg}.
\section{\label{app2}Non-$\cal{PT}$-symmetric oscillator}
\par
\begin{figure}[htb!]
\begin{center}
   \includegraphics[width=8cm]{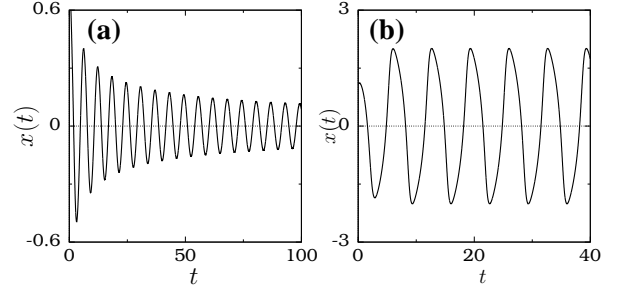}
\end{center}
   \caption{Temporal behavior of (a) non-$\cal{PT}$-symmetric damped oscillator Eq. (\ref{npt}) and (b) the limit cycle oscillator Eq. (\ref{lim}).}
\label{npt2}
 \end{figure}

 The non-$\cal{PT}$-symmetric oscillator given in Eq. (\ref{npt}), shows damped oscillations as given in Fig. \ref{npt2}(a). But the linear stability analysis of this system indicates a different dynamical behavior.  Note that this system has the same equilibrium points as that of (\ref{pt22}).  The eigenvalues associated with the equilibrium point $E_0$, namely $\pm i \sqrt{\lambda}$, show that it has periodic oscillations.  But the numerical results show that it has damped oscillations. The apparent ambiguity can be removed using its amplitude equation. We assume
\begin{eqnarray}
x(t)=R(t) e^{i \omega_0 t}+ R^*(t) e^{-i \omega_0 t}
\label{rt}
\end{eqnarray}
where $R(t)=r(t) e^{i\delta(t)}$, $r(t)$ and $\delta(t)$ are slowly varying amplitude and phase. By differentiating we have
\begin{eqnarray}
\dot{x}(t)&=&(\dot{R}(t) + i \omega_0 R(t)) e^{i \omega_0 t}+c.c.,\nonumber \\
\ddot{x}(t)&=&(\ddot{R}(t)+ 2 i \omega_0 \dot{R}(t) -\omega_0^2 R(t)) e^{i \omega_0 t}+c.c.,
\label{rt2}
\end{eqnarray}
where $c.c.$ denotes complex conjugate.  As $R(t)$ is a slowly varying quantity, $\dot{R}(t)<<\omega_0 R(t)$ and $\ddot{R}(t)<<\omega_0^2 \dot{R}(t)$.  Thus, we use approximations like 
\begin{eqnarray}
\dot{x}(t)&=& i \omega_0 R(t) e^{i \omega_0 t}+c.c., \nonumber \\
\ddot{x}(t)&=&( 2 i \omega_0 \dot{R}(t) -\omega_0^2 R(t)) e^{i \omega_0 t}+c.c.
\label{rt3}
\end{eqnarray}
Substituting (\ref{rt3}) and (\ref{rt}) in (\ref{npt}), we get for the equation for amplitude ($r(t)$),
\begin{eqnarray}
\dot{r}=-\alpha \frac{r^3(t)}{2}.
\label{eqr}
\end{eqnarray}
By solving the above, we get
\begin{eqnarray}
r(t)=\frac{1}{\sqrt{\alpha (t-t_0)}},\;\;   t_0, \mathrm{constant} 
\label{roft}
\end{eqnarray}
This indicates that the amplitude of oscillation decreases due to the introduced nonlinear term. This is the reason why the system in (\ref{npt}) has damped oscillations. 
\par On the other hand, the amplitude equation associated with $E_0$ corresponding to MEE (\ref{pt22}) is found to be 
\begin{eqnarray}
\dot{r}=0.
\label{eqr2}
\end{eqnarray}
Thus, $r(t)=$ constant, in the case of MEE. Thus, it has periodic oscillations with constant amplitude.
\par  Now considering the non-$\cal{PT}$-symmetric limit cycle oscillator equation given in (\ref{lim}), we see that it has an equilibrium point $E_0$ at $(0,0)$.  The associated eigenvalues are $\frac{1\pm \sqrt{1-4 \omega_0^2}}{2}$.  This shows that the system is unstable.  But the amplitude equation of the system (obtained as in the previous case) is
\begin{eqnarray}
\dot{r}=-\frac{r^3-r}{2}
\end{eqnarray}
indicates that $\dot{r}=0$ for $r=1$.  Thus the system exhibits limit cycle oscillations.
  Fig. \ref{npt2}(b) shows the limit cycle oscillation of (\ref{lim}).  
  
\section{\label{apeg1} Eigenvalues of Eq. (\ref{bip})}
In this section, we present the eigenvalues of the Jacobian matrix $J$ (given in (\ref{jac1})) associated with the various equilibrium points.  The eigenvalues of $J$ corresponding to the system (\ref{bip}) for the equilibrium point $e_0$ are
\begin{eqnarray}
\mu_{j}^{(0)} = \pm i \sqrt{\omega_0^2 \pm \kappa}, \qquad j=1,2,3,4.
\label{e0bip}
\end{eqnarray} 
The eigenvalues are found to be pure imaginary when $\omega_0^2 \leq \kappa \leq -\omega_0^2$.  In this range, for all values of the nonlinear damping coefficient $\alpha$, the eigenvalues are pure imaginary.  This indicates that there is no symmetry breaking while increasing $\alpha$. \\ Now, we consider the equilibrium points $e_{1,2}$ which exist only for $\kappa > \omega_0^2$. The eigenvalues of (\ref{jac1}) corresponding to $e_1$ and $e_2$ are the same and they are given by
\begin{small}
\begin{eqnarray}
\mu_{j}^{(1,2)}= \pm \sqrt{ \frac{b_{1} \pm \sqrt{b_{1}^2-b_2}}{2} }; \; j=1,2,3,4, \label{e1bip} 
\end{eqnarray}
\end{small}
where,
\begin{small}
\begin{eqnarray}
b_1 &=& \left(\alpha \sqrt{\frac{\kappa-\omega_0^2}{\beta}} \right)^2-(6 \kappa - 4 \omega_0^2), \label{b1bip}\\
b_2&=&16 (2 \kappa -\omega_0^2)(\kappa -\omega_0^2). \label{b2bip}
\end{eqnarray}
\end{small}
For fixed values of $\alpha$ and $\beta$, these eigenvalues are found to be pure imaginary for the values of $\kappa$ in the range
\begin{small}
\begin{eqnarray}
\omega_0^2<\kappa \leq \frac{(\alpha^2-6 \beta)(\alpha^2-4 \beta)-24 \beta^2- 4 \alpha \beta \sqrt{2 \beta}}{((\alpha^2-6\beta)^2-32 \beta^2)}\omega_0^2.
\label{kapp1}
\end{eqnarray}
\end{small}

Similarly, for a particular value of $\kappa$ in the range $\kappa> \omega_0^2$, the range of values of $\alpha$ for which the eigenvalues will be pure imaginary is given below,
\begin{small}
\begin{eqnarray}
-\sqrt{\frac{\beta}{\kappa-\omega_0^2}} \,b_3 \leq \alpha \leq \sqrt{\frac{\beta}{\kappa-\omega_0^2}}\, b_3 \label{c1bip}
\end{eqnarray}
\end{small}
where
\begin{small}
\begin{eqnarray}
b_3=\sqrt{6\kappa-4\omega_0^2-\sqrt{b_2}} \label{b3bip}
\end{eqnarray}
\end{small}
with the values of $\kappa \geq \omega_0^2$. 
\\ The eigenvalues of $J$ corresponding to the equilibrium point $e_{3}$ (which exists when $\kappa < \omega_0^2$) are
\begin{small}
\begin{eqnarray}
\mu_{1,2}^{(3)}=\frac{-\alpha\sqrt{-(\kappa+\omega_0^2)}\pm \sqrt{(-\alpha^2+8\beta)(\kappa+\omega_0^2)}}{2\sqrt{\beta}}\qquad \qquad \nonumber \\
\mu_{3,4}^{(3)}=\frac{-\alpha\sqrt{-(\kappa+\omega_0^2)}\pm \sqrt{-\alpha^2(\kappa+\omega_0^2)+8\beta(2 \kappa+\omega_0^2)}}{2\sqrt{\beta}}\quad
\label{e3bip}
\end{eqnarray}
\end{small}
We can find from the above equation that these eigenvalues can never be pure imaginary if $\alpha \neq 0$.  The equilibrium point $e_3$ is found to be stable and gives rise to oscillation death when $\alpha>0$.  The eigenvalues of $J$ corresponding to $e_4$ can be obtained by simply changing $\alpha \rightarrow -\alpha$ in Eq. (\ref{e3bip}). One can check that its eigenvalues can never be pure imaginary for $\alpha \neq 0$ and that they can become stable when $\alpha <0$.
\section{\label{apeg2} Eigenvalues of Eq. (\ref{bip2})}
In this appendix, we present the eigenvalues of $J$ given in (\ref{jac2}) for the equilibrium points of the system (\ref{bip2}).  This system has the same set of equilibrium points as that of (\ref{bip}).  The eigenvalues of $J$ for $e_0$ are
\begin{small}
\begin{eqnarray}
{\mu}^{(0)}_{j}=\pm \sqrt{\frac{-(2\omega_0^2-\gamma^2)\pm\sqrt{(2\omega_0^2-\gamma^2)^2+4(\kappa^2-\omega_0^4)}}{2}}.
\label{e0b2}
\end{eqnarray}
\end{small}
For the values of $\gamma$ in the range $-\sqrt{2 \omega^2_0}< \gamma< \sqrt{2 \omega_0^2}$, the eigenvalues are easily seen to be pure imaginary only for the values of $\kappa$ in the range 
\begin{eqnarray}
-\omega_0^2 < \kappa \leq -\sqrt{\frac{4 \omega_0^4-(2 \omega_0^2-\gamma^2)^2}{4}} \quad \mathrm{and} \nonumber \\
\omega_0^2 > \kappa \geq \sqrt{\frac{4 \omega_0^4-(2 \omega_0^2-\gamma^2)^2}{4}}. \qquad \quad
\label{kapb2}
\end{eqnarray}

 For a particular values of $\kappa$ in the region $-\omega_0^2\leq\kappa\leq\omega_0^2$, $e_0$ is neutrally stable for the values of $\gamma$ defined by
\begin{small}
\begin{eqnarray}
-\sqrt{2\omega_0^2-2\sqrt{\omega_0^4-\kappa^2}}\leq \gamma \leq \sqrt{2\omega_0^2-2\sqrt{\omega_0^4-\kappa^2}}.
\label{add}
\end{eqnarray}
\end{small}
  From the above relations, one can see that the increase in $\gamma$ beyond  this range causes symmetry breaking in the system (in the region  $-\omega_0^2 \leq \kappa \leq \omega_0^2$).
\par Then, the eigenvalues of $J$ for the equilibrium point $e_{1}$ are
\begin{small}
\begin{eqnarray}
{\mu}^{(1)}_{j}=\pm \sqrt{\frac{{b'}_1\pm \sqrt{({b'}_1^2-b_2)}}{2}} 
\label{e1b2}
\end{eqnarray}
\end{small}
where
\begin{small}
\begin{eqnarray}
{b'}_{1} &=& \left( \alpha \sqrt{\frac{\kappa-\omega_0^2}{\beta}}+\gamma \right)^2-(6 \kappa - 4 \omega_0^2), \label{b1b2}
\end{eqnarray}
\end{small}
and $b_2$ is given in (\ref{b2bip}).  The equilibrium point $e_1$ exists only when $\kappa > \omega_0^2$, and the associated eigenvalues are pure imaginary when
\begin{small}
\begin{eqnarray}
- \alpha\sqrt{\frac{\kappa-\omega_0^2}{\beta}}-b_3 \leq \gamma \leq  - \alpha\sqrt{\frac{\kappa-\omega_0^2}{\beta}}+b_3
\label{c2b2}
\end{eqnarray}
\end{small}
where $b_3$ is given in (\ref{b3bip}).
Thus, the $\cal{PT}$ symmetry is unbroken in the region given above.  Similarly, the eigenvalues of $J$ with respect to $e_2$ and the regions in which they take pure imaginary eigenvalues can be obtained by replacing $\alpha$ be $-\alpha$ in (\ref{e1b2}) and (\ref{c2b2}).\\
 Then, considering the equilibrium point $e_{3}$ (which exist for $\kappa \leq -\omega_0^2$), its eigenvalues are the roots of the algebraic equation
\begin{small}
\begin{eqnarray}
&&{{\mu}^{(3)}}^4 + 2 \alpha \sqrt{\frac{-(\kappa+\omega_0^2)}{\beta}}{{\mu}^{(3)}}^3 +(-\alpha^2\frac{(\kappa+\omega_0^2)}{\beta}-\gamma^2 \nonumber \\
&&-(6\kappa+4\omega_0^2)){{\mu}^{(3)}}^2+\alpha \sqrt{-\frac{(\kappa+\omega_0^2)}{\beta}}(6\kappa+4\omega_0^2){\mu}^{(3)} \nonumber \\
&& \qquad \qquad\qquad\qquad \; + 4(2 \kappa-\omega_0^2)(\kappa-\omega_0^2)=0.
\label{e34b2}
\end{eqnarray}  
\end{small}
As the coefficients of ${{\mu}^{(3)}}^3$ and ${{\mu}^{(3)}}$ are non-zero for $\alpha \neq 0$, $\beta \neq 0$, the eigenvalues of the equilibrium point cannot take pure imaginary values.  Similarly, the eigenvalue equation corresponding to the equilibrium point $e_4$ can be obtained by changing $\alpha \rightarrow -\alpha$ in (\ref{e34b2}).  
\section{\label{new} Eigenvalues of Eq. (\ref{bip3})}
Now we consider the general case of Eq. (\ref{bip3}), where we can choose $f(x)$ to be an odd or an even function.  In this section, depending on the nature of $f(x)$ (odd or even), we have presented their corresponding eigenvalues.
\subsection{\label{new1}Case: $f(x)$ - odd}
 In this case, the eigenvalues of $J$ corresponding to the equilibrium point $e_0$ are found to be the same as in (\ref{e0b2}). 
\par The eigenvalues about the equilibrium point $e_1$ and $e_2$ are 
\begin{small}
\begin{eqnarray}
{\mu}^{(1,2)}_{j}=\pm \sqrt{\frac{{\tilde{b}}_1^{(1,2)}\pm \sqrt{(({\tilde{b}}_1^{(1,2)})^2-b_2)}}{2}} 
\label{e1bg}
\end{eqnarray}
\end{small}
where
\begin{small}
\begin{eqnarray}
{\tilde{b}}^{(1)}_{1} &=& \left(+ \alpha f\left(\sqrt{\frac{\kappa-\omega_0^2}{\beta}}\right)+\gamma \right)^2-(6 \kappa - 4 \omega_0^2), \nonumber \\ 
{\tilde{b}}^{(2)}_{1} &=& \left(- \alpha f\left(\sqrt{\frac{\kappa-\omega_0^2}{\beta}}\right)+\gamma \right)^2-(6 \kappa - 4 \omega_0^2),\label{b1b2}
\end{eqnarray}
\end{small}
and $b_2$ is as given in (\ref{b2bip}). The regions in which the eigenvalues of $e_1$ and $e_2$ are found to be pure imaginary are given respectively by
\begin{small}
\begin{eqnarray}
- \alpha f\left(\sqrt{\frac{\kappa-\omega_0^2}{\beta}}\right)-b_3 \leq \gamma \leq  - \alpha f\left(\sqrt{\frac{\kappa-\omega_0^2}{\beta}}\right)+b_3 \;\; \mathrm{and} \nonumber \\
+ \alpha f\left(\sqrt{\frac{\kappa-\omega_0^2}{\beta}}\right)-b_3 \leq \gamma \leq  + \alpha f\left(\sqrt{\frac{\kappa-\omega_0^2}{\beta}}\right)+b_3.
\label{c2bg}
\end{eqnarray}
\end{small}
where $b_3$ is given in (\ref{b3bip}).

\subsection{\label{new2}Case: $f(x)$ - even}
\par Considering the case of even $f(x)$, the eigenvalues of $e_0$ are
\begin{small}
\begin{eqnarray}
{\mu}^{(0)}_{j}=\pm\sqrt{\frac{-c \pm \sqrt{c^2-4(\omega_0^4-\kappa^2)}}{2}}.
\label{e0bg2}
\end{eqnarray}
\end{small}
where $c=(2\omega_0^2-(\gamma-\alpha f(0))^2)$.  The eigenvalues in (\ref{e0bg2}) are found to be same as that of (\ref{e0b2}) when $f(0)=0$. In the case $f(0)=1$, thus the eigenvalues given in (\ref{e0bg2}) are different from that of (\ref{e0b2}). In contrast to the previous cases (Eq. (\ref{e0bip}) and (\ref{e0b2})), the eigenvalues corresponding to $e_0$ are found to depend on $\alpha$ and the region in which the eigenvalues given in (\ref{e0bg2}) take pure imaginary values is
\begin{small}
\begin{eqnarray}
\alpha-\sqrt{2\omega_0^2-\sqrt{4(\omega_0^4-\kappa^2)}} \leq \gamma  \leq \alpha+\sqrt{2\omega_0^2-\sqrt{4(\omega_0^4-\kappa^2)}}. \quad
\label{c0bg2}
\end{eqnarray}
\end{small}
The eigenvalues corresponding to both $e_1$ and $e_2$ are found to be the same and they are
\begin{small}
\begin{eqnarray}
{\mu}^{(1,2)}_{j}=- \sqrt{\frac{{\tilde{b}}_1^{(2)}\pm \sqrt{({\tilde{b}}_1^{(2)})^2-b_2}}{2}}.
\label{e1bg2}
\end{eqnarray}
\end{small}
The eigenvalues corresponding to both $e_1$ and $e_2$ are found to be pure imaginary only when
\begin{small}
\begin{eqnarray}
 \alpha f\left(\sqrt{\frac{\kappa-\omega_0^2}{\beta}}\right)-b_3 \leq \gamma  
\leq   \alpha f\left(\sqrt{\frac{\kappa-\omega_0^2}{\beta}}\right)+b_3.
\label{c3bg2}
\end{eqnarray}
\end{small}

\end{document}